\documentstyle[graphicx,prd,aps]{revtex}
\begin{document}


\title{Geodetic Brane Gravity}

\author{David Karasik and Aharon Davidson}
\address{Physics Department, Ben-Gurion University of the
Negev, Beer-Sheva 84105, Israel \\
({\sf karasik@bgumail.bgu.ac.il, davidson@bgumail.bgu.ac.il})}
\date{14.7.2002}

\maketitle

\begin{abstract}
Within the framework of geodetic brane gravity, 
the Universe is described as a 4-dimensional 
extended object evolving geodetically
in a higher dimensional flat background.
In this paper, by introducing a new pair of canonical fields
$\{\lambda, P_{\lambda}\}$, we derive the \textit{quadratic}
Hamiltonian for such a brane Universe; the inclusion of matter
then resembles minimal coupling.
Second class constraints enter the game, 
invoking the Dirac bracket formalism.
The algebra of the first class 
constraints is calculated, and the BRST
generator of the brane Universe turns out to be rank-$1$.
At the quantum level, the road is 
open for canonical and/or functional
integral quantization.
The main advantages of geodetic brane gravity are:
(i) It introduces an intrinsic, geometrically originated, 
'dark matter' component,
(ii) It offers, owing to the Lorentzian bulk 
time coordinate, a novel solution to the 
'problem of time', and
(iii) It enables calculation of meaningful 
probabilities within quantum
cosmology without any auxiliary scalar field.
Intriguingly, the general relativity limit 
is associated with $\lambda$
being a vanishing (degenerate) eigenvalue.
\end{abstract}



\section{Introduction}
Geodetic Brane Gravity (GBG) treats the universe
as an extended object (brane)
evolving geodetically in some flat background.
This idea has been proposed more than twenty
years ago by Regge and Teitelboim 
('General Relativity a la String') 
\cite{RT}, with the motivation that the first
principles which govern the evolution of the entire
universe cannot be too different from those which determine
the world-line behavior of a point particle or the
world-sheet behavior of a string.

Geometrically speaking, the $4$-dimensional curved
space-time is a hypersurface embedded within a higher 
dimensional flat manifold.
Following the isometric embedding theorems \cite{Cartan},
at most $N=\frac{1}{2}n(n+1)$ background flat dimensions
are required to \textit{locally} embed a general $n$-metric.
In particular, for $n=4$, one needs at most a $10$ dimensional
flat background. This number can be reduced, however, if the
$n$-metric admits some Killing-vector fields.

In Regge and Teitelboim (RT) model, 
the external manifold (the bulk) is flat
and empty, it contains neither a gravitational field 
nor matter fields. Other models were suggested, where
the external manifold is more complicated 
\cite{RS,Dvali,Ida,carter,Stealth},
it may be curved and contain bulk fields which 
may interact with the brane. 
RT action, therefore, does not contain bulk integrals,
it is only an integral over the brane manifold,
which may include the scalar curvature (${\cal R}^{n}$),
a constant ($\Lambda$), and some matter Lagrangian 
(${\cal L}_{matter}$)
\footnote{ The $n=1$ brane is a particle, it has ${\cal R}^{1}=0$ 
and $\Lambda$ is the mass of the particle. 
The $n=2$ brane is a string, it's curvature ${\cal R}^{2}$ is 
just a topological term, and $\Lambda$ is the string tension. 
The brane universe $n=4$ includes both the scalar 
curvature ${\cal R}^{4}$, and the cosmological constant $\Lambda$. 
}.
\begin{equation}
    S = \int\left(\frac{1}{16\pi G^{n}}{\cal R}^{n}+\Lambda 
    + {\cal L}_{matter}\right)\sqrt{-g^{n}}\,d^{n-1}x\,d\tau 
    \label{action}
\end{equation}
The Geodetic Brane has two parents:
\begin{enumerate}
\item General relativity gave the Einstein-Hilbert action,
which makes the geodetic brane a gravitational theory.
\item Particle/String theory gave the embedding coordinates 
$y^{A}(x)$ 
\footnote{We denote the embedding space indices with upper-case 
Latin letters, spacetime indices with Greek letters, 
and space indices with lower-case Latin letters. 
$\eta_{AB}$ is the Minkowski metric of the embedding space.}
as canonical fields, and this will lead to geodetic evolution. 
The $4$ dimensional metric is not a canonical field,
it is just being induced by the embedding 
$g_{\mu\nu}(x)=\eta_{AB}y^{A}_{\,,\mu}(x)y^{B}_{\,,\nu}(x)$.
\end{enumerate}
Due to the fact that the Lagrangian (\ref{action})
does not depend explicitly on $y^{A}$, but solely 
on the derivatives through the metric,
the geodetic brane equations of motion are actually 
a set of conservation laws 
\begin{equation}
	\left[({\cal R}^{\mu\nu}-\frac{1}{2}g^{\mu\nu}{\cal R} -8\pi
	GT^{\mu\nu})y^{A}_{\,;\mu}\right]_{;\nu}=0.
	\label{RT1}
\end{equation} 
Eq.(\ref{RT1}) splits into two parts, the first is proportional
to $y^{A}_{\,,\mu}$ and the second to $y^{A}_{\,;\mu\nu}$.
Since the $4$-dimensional covariant 
derivative of the metric vanishes 
$g_{\mu\nu;\lambda}=0$, one faces the embedding identity
$\eta_{AB}y^{A}_{\,;\lambda}y^{B}_{\,;\mu\nu}=0$.
Therefore, the first and second covariant derivatives 
of $y^{A}$, viewed as vectors in the external manifold, are 
orthogonal, and each part of Eq.(\ref{RT1}) should vanish separately.
The part proportional to $y^{A}_{\,,\mu}$ implies that
$T^{\mu\nu}_{\,;\nu}=0$. 
The second part is the geodetic brane equation 
\footnote{The geodetic factor 
$y^{A}_{\,;\mu\nu}-\Gamma^{A}_{BC}y^{B}_{\,,\mu}y^{C}_{\,,\nu}$ 
replaces $y^{A}_{\,;\mu\nu}$ in case 
the embedding metric is not Minkowski.}
\begin{equation}
    \left({\cal R}^{\mu\nu} - \frac{1}{2}g^{\mu\nu}
    {\cal R} - 8\pi GT^{\mu\nu}\right)y^{A}_{\,;\mu\nu}
    = 0 ~.
    \label{RTeq}
\end{equation}
\begin{itemize}
\item The matter fields equations remain intact, since 
the matter Lagrangian depends only on the metric.
\item Energy momentum is conserved. This is a crucial result,
especially when the Einstein equations are not at our disposal.
\item Clearly, every solution of Einstein equations is
automatically a solution of the corresponding geodetic brane equations.
But the geodetic brane equations allow for different 
solutions \cite{Deser}.
A general solution of eq.(\ref{RTeq}) may look like
\begin{mathletters}
\begin{eqnarray}
    & & {\cal R}^{\mu\nu} - \frac{1}{2}g^{\mu\nu}
    {\cal R} - 8\pi GT^{\mu\nu}=D^{\mu\nu} \label{RT2} \\
    & & D^{\mu\nu}y^{A}_{\,;\mu\nu} = 0 \;\;\; D^{\mu\nu}\neq 0
\end{eqnarray}
\end{mathletters}
The non vanishing right hand side of eq.(\ref{RT2}) will 
be interpreted by an {\em Einstein Physicist} as additional matter,
and since it is not the ordinary $T^{\mu\nu}$ it may labeled
{\em Dark Matter} \cite{DKL}.
\end{itemize}
It has been speculated, relying on the structural similarity
to string/membrane theory, 
that quantum geodetic brane gravity may be a
somewhat easier task to achieve 
than quantum general relativity (GR).
The trouble is, however, that
the parent Regge-Teitelboim \cite{RT} 
Hamiltonian has never been derived!

In this paper, by adding a new non-dynamical canonical field
$\lambda$ we derive the quadratic Hamiltonian density of a
gravitating brane-universe
\begin{equation}
    {\cal H}=N^{k}y_{|k}\cdot P
    - N\frac{8\pi G}{2\sqrt{h}}\left[
    \left(\frac{\sqrt{h}}{8\pi G}\right)^{2}(\lambda + {\cal R}^{(3)})
    + P\Theta(\Psi - \lambda I)^{-1}\Theta P \right]  ~.
\end{equation}

The derivation of the Geodetic Brane Hamiltonian 
is done here in a pedagogical way.
In section \ref{embedding} we translate the relevant 
geometric objects to the language
of embedding. Each object is 
characterized by its tensorial 
properties with respect to both the 
embedding manifold and the brane manifold. 
We embed the ADM formalism \cite{ADM} in a higher dimensional Minkowski
background, the $4$ dimensional spacetime 
manifold ($V_{4}$)is artificially separated into a
$3$-dimensional space-like manifold ($V_{3}$) 
and a time direction characterized by
the time-like unit vector orthogonal to $V_{3}$. 
For simplicity we restrict ourselves to $3$-dimensional 
space-like manifolds with no boundary
(either compact or infinite), while the
appropriate surface terms should be added when 
boundaries are present \cite{boundary}. 

Section \ref{hamiltonian} is the main part of this
Paper, where we derive the Hamiltonian.
We first look at an empty 
universe with no matter fields, 
we present the gravitational 
Lagrangian density as a functional of the embedding
vector $y^{A}(x)$, and derive the conjugate momenta $P_{A}(x)$. 
Reparametrization invariance causes the
canonical Hamiltonian to vanish, (in a similar way to the
ADM-Hamiltonian and string theory), and the total Hamiltonian is a
sum of constraints. We introduce a
new pair of canonical fields $\lambda,P_{\lambda}$ and make the 
Hamiltonian quadratic in the momenta.
Following Dirac's procedure \cite{Dirac} we separate the
constraints into $4$ first-class constraints (reflecting
reparametrization invariance), and $2$ second-class constraints
(caused by the $2$ extra fields). We define the Dirac-Brackets
and eliminate the second-class constraints. The final algebra of
the constraints takes the familiar form of a relativistic theory,
such as: The relativistic particle, string or membrane.

In section \ref{matter} we discuss 
the inclusion of arbitrary matter fields confined to the 
four dimensional brane. 
The algebra of the constraints
remains unchanged, while the Hamiltonian is simply the sum
of the gravitational Hamiltonian and the matter Hamiltonian.

In section \ref{einstein} the necessary 
conditions for classical 
Einstein gravity are formulated, they are
\begin{itemize}
    \item $\lambda$ must vanish.
    \item The total (bulk) momentum of the brane vanishes.
\end{itemize}

Section \ref{quantization} deals with quantization schemes.
We can use canonical quantization by setting 
the Dirac Brackets to be commutators
$\displaystyle{\left\{,\right\}_{D} \longrightarrow i\hbar
\left[,\right]}$. The wave-functional of a brane-like Universe
\cite{HH} is subject to a Virasoro-type momentum constraint
equation followed by a Wheeler-deWitt-like equation (first class
constraints), the operators are not free, but are constrained by
the second class constraints as operator identities. 
Another quantization scheme is the functional integral formalism, 
where we use 
the BFV \cite{BFV} formulation. The BRST generator \cite{BRST}
is calculated, and the theory turns out to be rank $1$.
This resembles ordinary gravity and string theory as oppose to 
membrane theory, where the rank is the dimension of the 
underlying space manifold.

Section \ref{mini} Geodetic Brane Quantum Cosmology is demonstrated.
We apply the path integral quantization to the homogeneous
and isotropic geodetic brane, within the minisuperspace model. 
A possible solution to the 
problem of time arises when one notices that while in GR the 
only dynamical degree of freedom is the scale factor of the universe, 
GBQC offers one extra dynamical degree of freedom (the bulk time)
that may serve as time coordinate.

Definitions, notations and some lengthy calculations
were removed from the main stream of this work and 
were put in the appendix section.  

\section{The Geometry of Embedding}
\label{embedding}
In this section we will formulate the relevant geometrical 
objects of the $V_{4}$ and $V_{3}$ manifolds in the language
of embedding.
Let our starting point be a flat $m$-dimensional manifold ${\cal M}$, 
with the corresponding line-element being
\begin{equation}
    ds^{2}=\eta_{AB}dy^{A}dy^{B} ~.
\end{equation}
\begin{itemize}
\item\underline{Hypersurfaces :}
An embedding function $y^{A}(x^{\mu}) \; (\mu=0,1,2,3)$ defines the 
$4$ dimensional hypersurface $V_{4}$ parameterized by the $4$ 
coordinates $x^{\mu}$. The $V_{4}$ tangent 
space is spanned by the vectors $y^{A}_{\,,\mu}$. 
(The $V_{3}$ hypersurface and tangent space are defined 
in a similar way).
The induced $4$-dimensional metric is the projection of $\eta_{AB}$
onto the $V_{4}$ manifold:
$g_{\mu\nu}=\eta_{AB}y^{A}_{\,,\mu}y^{B}_{\,,\nu}$.
Choosing a time direction $t$ and 
space coordinates $x^{i} \; (i=1,2,3)$,
the induced $4$-dimensional line-element takes the form
\begin{equation}
    ds^{2} = \eta_{AB}(y^{A}_{\, ,i}dx^{i}+
    \dot{y}^{A}dt)(y^{B}_{\,,j}dx^{j}+
    \dot{y}^{B}dt) ~,
\end{equation}

The various projections of the metric $\eta_{AB}$ onto the
space and time directions are denoted as
the $3$-metric $h_{ij}$, the shift vector $N_{i}$, and
the lapse function $N$
\begin{mathletters}
\begin{eqnarray}
    \eta_{AB}y^{A}_{\,,i}y^{B}_{\,,j} & = & h_{ij} ~,
    \label{ADMa} \\
    \eta_{AB}y^{A}_{\,,i}\dot{y}^{B} & = & N_{i} ~,
    \label{ADMb} \\
    \eta_{AB}\dot{y}^{A}\dot{y}^{B} & = &
    N_{i}N^{i} - N^{2} ~.
    \label{ADMc}
\end{eqnarray}
\end{mathletters}
These are not independent fields (as in Einstein's gravity),
but are functions of the embedding vector $y^{A}$. 
Nevertheless, it is a matter of convenience to write down 
the induced $4$-dimensional line-element in
the familiar Arnowitt-Deser-Misner \cite{ADM} (ADM) form
\begin{equation}
    ds^{2}= -N^{2}dt^{2}+
    h_{ij}(dx^{i}+N^{i}dt)(dx^{j}+N^{j}dt) ~.
\end{equation}
The vectors $(\dot{y}^{A},y^{A}_{\,,i})$ span the $4$-dimensional
tangent-space of the $V_{4}$ spacetime manifold, while
$y^{A}_{\,,i}$ span the $3$-dimensional tangent-space of the
$V_{3}$ manifold.
Using $h^{ij}$ as the inverse of the $3$-metric 
$h^{ij}h_{jk}=\delta^{i}_{k}$, one 
can introduce projections orthogonal to the $V_{3}$ manifold
with the operator 
\begin{mathletters}
\begin{eqnarray}
	\Theta^{A}_{B} & = & \delta^{A}_{B}- y^{A}_{\,,a}h^{ab} y_{B,b} ~.\\
	\Theta^{A}_{C}\Theta^{C}_{B} & = & \Theta^{A}_{B} ~.
	\label{Theta} 
\end{eqnarray}
\end{mathletters}
Now, any vector $v^{A}$ can be separated into the projections tangent
and orthogonal to the $V_{3}$ space
\begin{equation}
    v^{A}= v^{A}_{\parallel} + v^{A}_{\perp} = v^{B}y_{B,b}h^{ab}y^{A}_{\,,a}
    + v^{B}(\delta^{A}_{B}- y^{A}_{\,,a}h^{ab} y_{B,b}) ~.
\end{equation}
An important role is played by the time-like unit vector
orthogonal to $V_{3}$-space yet tangent to $V_{4}$-spacetime,
\begin{mathletters}
\begin{eqnarray}
    n^{A} & \equiv & \frac{1}{N}\left(\dot{y}^{A}-
    N^{i}y^{A}_{\,,i}\right)
    = \frac{1}{N}\dot{y}^{B}\Theta^{A}_{B} ~,
    \label{n} \\
    \eta_{AB}y^{A}_{\,,i}n^{B} & = & 0 ~, \\
    \eta_{AB}n^{A}n^{B} & = & -1 ~.
\end{eqnarray}
\end{mathletters}
The tangent space of the embedding manifold ${\cal M}$ is
spanned by the vectors: $y^{A}_{\,,i}$, $n^{A}$ and $L^{A}_{p}$ 
($i=1,2,3\;p=1,..,m-4$). The vectors $L^{A}_{p}$ are
chosen to be orthogonal to $y^{A}_{\,,i}$, $n^{A}$ and to each other.
\item\underline{Curvature :}
The connections on the underlying $V_{3}$ are 
$\Gamma^{k}_{ij}=\eta_{AB}y^{A}_{\,,ij}y^{B}_{\,,l}h^{kl}$,
this way, the covariant derivative of the $3$-metric
vanishes $h_{ij|k}=0$ (the stroke denotes 3-dimensional
covariant derivative). As a result, 
one faces the powerful \textit{embedding identity}
\begin{equation}
    \eta_{AB}y^{A}_{\,|ij}y^{B}_{\,,k} \equiv 0 ~.
    \label{RT}
\end{equation}
The vectors $y^{A}_{\,|ij}$ are 
orthogonal to the $V_{3}$ tangent space and may be written
as a combination of $n^{A}$ and $L^{A}_{p}$ \cite{exact}.
\begin{equation}
    y^{A}_{\,|ij}=n^{A}K_{ij} + L^{A}_{p}\Omega^{p}_{ij} ~.
\end{equation}
The projection of $y^{A}_{\,|ij}$ in the $n^{A}$ direction
is the extrinsic curvature of the $V_{3}$
hypersurface embedded in $V_{4}$
\begin{equation}
    K_{ij} \equiv -\frac{1}{2N}\left( N_{i|j}+N_{j|i}
    -\frac{\partial h_{ij}}{\partial t}\right)=
    -\eta_{AB}y^{A}_{\,|ij}n^{B} ~.
    \label{excurv}
\end{equation}
The coefficient $\Omega^{p}_{ij}$ is the extrinsic curvature 
of $V_{3}$ with respect to the corresponding 
normal vector $L^{A}_{p}$.
 
The intrinsic curvature of the $V_{3}$ manifold is also related to
the second derivative of the embedding functions $y^{A}_{\,|ij}$.
The $3$-dimensional Riemann tensor is
\begin{equation}
    {\cal R}^{(3)}_{iljk} \equiv \eta_{AB}(y^{A}_{\,|ij}y^{B}_{\,|kl}
        - y^{A}_{\,|ik}y^{B}_{\,|jl}) ~.
    \label{Rim}
\end{equation}
For convenience we define the $\dot{y}^{A}$-independent symmetric
tensor
\begin{equation}
    \Psi^{AB} \equiv (h^{ij}h^{ab}-h^{ia}h^{jb})
    y^{A}_{\,|ij}y^{B}_{\,|ab} ~.
    \label{Psi}
\end{equation}
Checking the indices, $\Psi^{AB}$ is a tensor in the embedding
manifold, but a scalar in $V_{3}$ space.
The trace of $\Psi^{A}_{\,B}$ is simply the $3$-dimensional 
Ricci scalar
${\cal R}^{(3)} = \eta_{AB}\Psi^{AB}$.
Looking at eq.(\ref{RT}), one can easily check that
\begin{equation}
    \Psi^{A}_{\,B}y^{B}_{\,,i} = 0 ~,
    \label{Psiy}
\end{equation}
and $\Psi$ as an operator has at least $3$ eigenvectors
with vanishing eigenvalue.
Using definitions (\ref{Psi},\ref{excurv}), the contraction of
$\Psi$ twice with $n^{A}$ is related to the extrinsic curvature
\begin{equation}
    K^{i}_{\,i}K^{j}_{\,j} - K_{ij}K^{ij} =
    \Psi_{AB}n^{A}n^{B} = \frac{1}{N^{2}}
    \Psi_{AB}\dot{y}^{A}\dot{y}^{B} ~.
\end{equation}
\end{itemize}


\section{Deriving the Hamiltonian}
\label{hamiltonian}
The gravitational Lagrangian density is the standard one
\begin{equation}
    {\cal L} = \frac{1}{16\pi G}\sqrt{-g}{\cal R}^{(4)} ~.
\end{equation}
Up to a surface term, it can be written in the form
\begin{equation}
    {\cal L} = \frac{1}{16\pi G}N\sqrt{h}\left[
    {\cal R}^{(3)} - ( K^{i}_{\,i}K^{j}_{\,j}
    - K_{ij}K^{ij} ) \right] ~.
    \label{lang}
\end{equation}
Here, ${\cal R}^{(3)}$ denotes the 3-dimensional Ricci scalar, 
constructed by means of the 3-metric $h_{ij}$ (\ref{ADMa}), 
whereas $K_{ij}$ (\ref{excurv}) is the
extrinsic curvature of $V_{3}$ embedded in $V_{4}$.
Using the tensor $\Psi^{AB}$ (\ref{Psi}) one can put
the Lagrangian density (\ref{lang}) in the form
\begin{equation}
    {\cal L} = \frac{\sqrt{h}}{16\pi G}\left[
    N{\cal R}^{(3)} - \frac{1}{N}\Psi_{AB}\dot{y}^{A}
    \dot{y}^{B} \right] ~.
    \label{lang2}
\end{equation}
As one can see, the Lagrangian (\ref{lang2}) does not involve mixed
derivative $\dot{y}^{A}_{\, ,i}$ or second time derivative
$\ddot{y}^{A}$. The first derivative $\dot{y}^{A}$ appears either 
explicitly or within $N$.
Therefore the Lagrangian
\begin{equation}
    \fbox{${\cal L}(y,\dot{y},y_{|i},y_{|ij})$}
\end{equation}
is ripe for the Hamiltonian formalism.

The momenta $P_{A}$ conjugate to $y^{A}$ is simply
\begin{equation}
    P_{A}(x) \equiv \frac{\delta L}{\delta
    \dot{y}^{A}(x)} = \frac{\sqrt{h}}{16\pi G}\left\{
    \left[{\cal R}^{(3)} + \frac{1}{N^{2}}\Psi_{BC}
    \dot{y}^{B}\dot{y}^{C}\right]\frac{\partial N}
    {\partial \dot{y}^{A}}-\frac{2}{N}
    \Psi_{AB}\dot{y}^{B}\right\} ~.
    \label{P1}
\end{equation}
Using eq.(\ref{ADMb},\ref{ADMc}) to get
$\frac{\partial N}{\partial \dot{y}^{A}} = -n_{A}$, while
eq.(\ref{Psiy}) tells us that
$\frac{1}{N}\Psi_{AB}\dot{y}^{B} = \Psi_{AB}n^{B}$,
the momentum (\ref{P1}) becomes
\begin{equation}
    P^{A} = -\frac{\sqrt{h}}{16\pi G}\left\{\left[
    {\cal R}^{(3)} + n_{B}\Psi^{BC}n_{C} \right]
    n^{A} + 2\Psi^{A}_{\,B}n^{B} \right\} ~.
    \label{P2}
\end{equation}
The next step should be :"Solve eq.(\ref{P2}) for
$\dot{y}^{A}(y,P,y_{|i},y_{|ij})$".
But eq.(\ref{P2}) involves only $n^{A}$, so one would like
to solve eq.(\ref{P2}) for $n^{A}(P,y,y_{|i},y_{|ij})$
first, and then to solve eq.(\ref{n}) for $\dot{y}^{A}$
\begin{equation}
    \dot{y}^{A} = Nn^{A} + N^{i}y^{A}_{\,,i} ~.
\end{equation}
This looks innocent but even if one is able to solve eq.(\ref{P2})
for $n^{A}$, any attempt to solve eq.(\ref{ADMb}) for
$N^{i}(n,y,y_{|i})$ and eq.(\ref{ADMc}) for $N(n,y,y_{|i})$ will
lead to a cyclic redefinition of $N^{i}$ and $N$. This situation
is similar to other reparametrization invariant theories (such as
the relativistic free particle, string theory etc.) and simply
means that we have here $4 \times V_{3}$ primary constraints
\begin{mathletters}
\begin{eqnarray}
    & & \eta_{AB}n^{A}n^{B} + 1 = 0  ~,
    \label{nn} \\
    & & \eta_{AB}y^{A}_{\,,i}n^{B} = 0 ~.
    \label{ny}
\end{eqnarray}
\end{mathletters}
The constraints should be written in terms of canonical fields
$(y^{A},P_{A})$. So one should solve eq.(\ref{P2}) for $n^{A}(P)$,
and then substitute in the above constraints.
Any naive attempt to solve eq.(\ref{P2}) for $n^{A}(y,P)$
falls short. The cubic equation involved rarely admits
simple solutions.
To 'linearize' the problem we define a new quantity
$\lambda$, such that
\begin{equation}
    P^{A}=-\frac{\sqrt{h}}{8\pi G}
    \left(\Psi - \lambda I\right)^{A}_{B}n^{B} ~.
    \label{P(n)}
\end{equation}
\begin{itemize}
\item   Comparing eq.(\ref{P2}) with eq.(\ref{P(n)}), the
    definition of $\lambda$ is actually another constraint
    \begin{equation}
        n_{A}\Psi^{A}_{\,B}n^{B} + {\cal R}^{(3)}
        + 2\lambda = 0 ~.
        \label{lambda}
    \end{equation}
\item   An \textit{independent} $\lambda$ comes along with its
    conjugate momentum $P_{\lambda}$. $\lambda$ is not
    a dynamical field therefore one faces another constraint
    \begin{equation}
        P_{\lambda} = 0 ~.
        \label{P_lambda}
    \end{equation}
\end{itemize}
Assuming $\lambda$ is \textit{\textbf{not}} an eigenvalue of
$\Psi^{A}_{\,B}$, we solve (\ref{P(n)}) for $n^{A}(\sqrt{h} ,\Psi
,P ,\lambda)$ and find
\begin{equation}
    n^{A}= -\frac{8\pi G}{\sqrt{h}}\left[\left(\Psi-\lambda
    I\right)^{-1}\right]^{A}_{\,B}P^{B} ~.
    \label{n(p)}
\end{equation}
At this point we have $6 \times V_{3}$ primary constraints
(\ref{nn},\ref{ny},\ref{lambda},\ref{P_lambda}). We will follow
Dirac's way \cite{Dirac} to treat the
\textit{constrained field theory} we have in hand.

First we will write down the various constraints in term
of the canonical fields $\left( y^{A}(x), P_{A}(x), \lambda(x),
P_{\lambda}(x)\right)$ :
\begin{mathletters}
\begin{eqnarray}
    \phi_{0} & = & \frac{8\pi G}{2\sqrt{h}}\left[
    \left(\frac{\sqrt{h}}{8\pi G}\right)^{2}(\lambda + {\cal R}^{(3)})
    + P\Theta(\Psi - \lambda I)^{-1}\Theta P 
    \right]\approx 0 ~,
    \label{phi0} \\
    \phi_{k} & = & y_{\,|k}\cdot P \approx 0 ~,
    \label{phik} \\
    \phi_{4} & = & P_{\lambda} \approx 0 ~,
    \label{phi4} \\
    \phi_{5} & = & \frac{8\pi G}{2\sqrt{h}}\left[
    \left(\frac{\sqrt{h}}{8\pi G}\right)^{2}
    + P\Theta(\Psi - \lambda I)^{-2}\Theta P 
    \right]\approx 0  ~.
    \label{phi5}
\end{eqnarray}
\label{phis}
\end{mathletters}
\underline{\textit{Notations :}}
\begin{itemize}
  \item We use shorthanded notation to simplify the detailed expressions,
  	$F \cdot G \equiv F^{A}G_{A}$ where $F$ and $G$ are vectors in the
  	embedding space, and $P(\Psi - \lambda I)^{-2}P \equiv
  	P_{A}\left[(\Psi - \lambda I)^{-2}\right]^{AB}P_{B}$
  \item We adopt Dirac's notation $\phi \approx 0$ for weakly
  	vanishing terms.
  \item The embedding functions $y^{A}(x)$ and $\lambda(x)$
  	are scalars in the $V_{3}$ manifold.
  	Their conjugate momenta $P_{A}(x),P_{\lambda}(x)$ are scalar densities
  	of weight $1$. For convenience we normalize all constraints 
	to be scalars in the
  	embedding space, and scalar/vector densities of weight $1$
  	in $V_{3}$. This way, the Lagrange multipliers are of weight $0$.
  \item $\phi_{k}$ is based on the constraint (\ref{ny}) but it
  	takes into account the embedding identity (\ref{Psiy})
  	  \begin{equation}
   	     \phi_{k} = y_{\,|k}\cdot P
  	      = -\frac{\sqrt{h}}{8\pi G }y_{\,|k}(\Psi - \lambda I)n
   	     = \frac{\lambda \sqrt{h}}{8\pi G }y_{\,|k}\cdot n \approx 0 ~.
  	  \end{equation}
  \item $\phi_{5}$ is based on the constraint (\ref{nn}), but
 	we added the projection operator $\Theta$ (\ref{Theta}) 
	in front of $P$.
 	This step simplifies the final algebra of the constraints,
 	and brings it to the familiar form of a relativistic theory.
	Inserting $\Theta$ in front of $P$, is equivalent to adding
	terms proportional to $\phi_{k}$ (\ref{phik}), since		
 	\begin{equation}
		\Theta^{A}_{B}P^{B}=(\delta^{A}_{B}
-y^{A}_{\,,a}h^{ab}y_{B,b})P^{B}
		=P^{A}-y^{A}_{\,,a}h^{ab}\phi_{b}~.
 	\end{equation}
 \item $\phi_{0}$ is also a combination of the constraints
 	 (\ref{lambda}),(\ref{ny}) and (\ref{nn}), chosen such that
 	\begin{equation}
 	   \frac{\partial\phi_{0}}{\partial\lambda}=
 	   \phi_{5} \approx 0 ~.
 	\end{equation}
 \item See appendix \ref{app FD} for the definitions of functional
 	 derivatives and Poisson brackets.
\end{itemize}
In a similar way to other parameterized theories, 
the canonical Hamiltonian density vanishes
\begin{equation}
    {\cal H}_{c} = \dot{y}^{A}P_{A}-{\cal L} \approx 0 ~.
\end{equation}
This means that the total Hamiltonian is a sum of
constraints
\begin{equation}
    H=\int d^{3}x\,u^{m}(x)\phi_{m}(x) ~.
    \label{H1}
\end{equation}
The constraints (\ref{phis}) should vanish for all times,
therefore their PB with the Hamiltonian should vanish (at least
weakly). This imposes a set of consistency conditions for the
functions $u^{m}(x)$
\begin{eqnarray}
    \dot\phi_{n}(x)=\left\{\phi_{n}(x),H\right\}
        & = & \left\{\phi_{n}(x),\int d^{3}z\,u^{m}(z)
    \phi_{m}(z)\right\} \nonumber \\
    & \approx & \int d^{3}z
    \,u^{m}(z)\left\{\phi_{n}(x),\phi_{m}(z)\right\}
        \approx 0 ~.
    \label{consis}
\end{eqnarray}
The basic Poisson brackets between the constraints are
calculated in appendix \ref{app DB}, and in general has the form
\begin{equation}
\begin{array}{|c|c|c|c|c|}
\hline
        \displaystyle{\left\{\,,\,\right\}\approx} & \phi_{0}(z) &
    \phi_{l}(z) & \phi_{4}(z) & \phi_{5}(z) \\
\hline
        \phi_{0}(x) & 0 & 0 & 0 & \frac{8\pi G}{2\sqrt{h}}\,\alpha(x,z) \\
\hline
        \phi_{k}(x) & 0 & 0 & 0 & \phi_{5,\lambda}\lambda_{|k}\delta(x-z) \\
\hline
        \phi_{4}(x) & 0 & 0 & 0 & -\phi_{5,\lambda}\delta(x-z) \\
\hline
        \phi_{5}(x) & -\frac{8\pi G}{2\sqrt{h}}\,\alpha(z,x) &
	 -\phi_{5,\lambda}\lambda_{|k}\delta(x-z) & 
	\phi_{5,\lambda}\delta(x-z) & [F^{i}(x)+F^{i}(z)]\delta_{|i}(x-z) \\
\hline
\end{array}
\label{PBarray}
\end{equation}
The exact expressions for $\alpha$ and $F^{i}$ appears in appendix \ref{app DB}.
Now, insert the PB between the constraints (\ref{PBarray}) into the
consistency conditions (\ref{consis}) to determine $u^{m}(x)$
\begin{eqnarray}
    \left\{\phi_{4}(x),H\right\} & \approx &
    \frac{\partial\phi_{5}}{\partial\lambda}(x)u^{5}(x)
    \approx 0 \Rightarrow u^{5}(x)=0 ~, \\
    \left\{\phi_{0}(x),H\right\} & \approx & 0
    \Rightarrow u^{0}(x) = -N(x)
     \; \emph{arbitrary} ~, \\
    \left\{\phi_{k}(x),H\right\} & \approx & 0
    \Rightarrow u^{k}(x) = N^{k}(x) \; \emph{arbitrary} ~, \\
    \left\{\phi_{5}(x),H\right\} & \approx & \int d^{3}z
    \left[\frac{8\pi G}{2\sqrt{h}}\,\alpha(z,x)N(z)
    -\phi_{5,\lambda}\lambda_{|k}\delta(x-z)N^{k}(z)
    +\phi_{5,\lambda}\delta(x-z)u^{4}(z)\right]
    \nonumber \\
    & & \Rightarrow u^{4}(x) = N^{k}\lambda_{|k}(x)-
 	\phi_{5,\lambda}^{-1}(x)
    \int d^{3}z\, {\text{ $\frac{8\pi G}{2\sqrt{h}}$}} \,\alpha(z,x)N(z) ~.
\end{eqnarray}
The first class Hamiltonian is then
\begin{eqnarray}
    H & = & \int d^{3}x\left\{\frac{}{}N^{k} \left[y_{|k}\cdot P
    + \lambda_{|k}P_{\lambda}\right] \right.\nonumber \\
    & -& N\frac{8\pi G}{2\sqrt{h}}\left[
    (\frac{\sqrt{h}}{8\pi G})^{2}(\lambda + {\cal R}^{(3)})
    + P\Theta(\Psi - \lambda I)^{-1}\Theta P 
	\right. \nonumber \\
    & & ~~~ \left. \left. +\int d^{3}z\,\alpha(x,z)
	\phi_{5,\lambda}^{-1}(z)P_{\lambda}(z)
    \right] \right\}
    \label{HT}
\end{eqnarray}
As one can see, at this stage we have in the
Hamiltonian $4$ arbitrary functions $N,N^{k}$ 
(Lagrange multipliers). This means we have $4$ first
class constraints reflecting the reparametrization invariance
($4$-dimensional general coordinate transformation)
\begin{mathletters}
\begin{eqnarray}
    \varphi_{0} & = & \frac{8\pi G}{2\sqrt{h}}\left[
    (\frac{\sqrt{h}}{8\pi G})^{2}(\lambda + {\cal R}^{(3)})
    + P\Theta(\Psi - \lambda I)^{-1}\Theta P 
	\right. \nonumber \\
    & & + \left. \int d^{3}z\,\alpha(x,z)
	\phi_{5,\lambda}^{-1}(z)P_{\lambda}(z)  \right] \approx 0 ~,
    \label{varphi0} \\
        \varphi_{k} & = & y_{\,|k}P
        + \lambda_{|k}P_{\lambda} \approx 0 ~.
        \label{varphik}
\end{eqnarray}
\label{varphi}
\end{mathletters}
And we are left with $2$ second class constraints, reflecting
the fact that we expanded our phase space with two
extra fields $\lambda$ and $P_{\lambda}$
\begin{mathletters}
\begin{eqnarray}
     \theta_{1}=\phi_{4} & = & P_{\lambda} \approx 0 ~,
    \label{theta1} \\
    \theta_{2}=\phi_{5} & = & \frac{8\pi G}{2\sqrt{h}}\left[
    \left(\frac{\sqrt{h}}{8\pi G}\right)^{2}
    + P\Theta(\Psi - \lambda I)^{-2}\Theta P 
	\right] \approx 0   ~.
    \label{theta2}
\end{eqnarray}
\end{mathletters}
Using the classical equation of motion for $y^{A}(x)$,
\begin{equation}
    \dot{y}^{A}(x) = \left\{y^{A}(x),H\right\} \approx
     N^{k}y_{|k} - N\frac{8\pi G}{\sqrt{h}}
    (\Psi - \lambda I)^{-1}P ~,
\end{equation}
one can identify 
the lapse function (\ref{ADMc}) and the shift vector
(\ref{ADMb}) with $N,N^{k}$ respectively. Thus, recover the
nature of the lapse function and the shift vector
as Lagrange multipliers only at the stage of the solution to the
equation of motion, not as an a priori definition.   

We would like to continue along Dirac's path \cite{Dirac}, and use
{\em{Dirac Brackets}} (DB) instead of Poisson Brackets (PB).
The DB are designed in a way such that the DB of a first class
constraint with anything is weakly the same as the corresponding PB,
while the DB of a second class constraint with anything vanish
identically. Using DB, we actually eliminate
the second class constraints (the extra degrees of freedom). 
The DB are defined as
\begin{equation}
    \left\{A,B\right\}_{D} \equiv \left\{A,B\right\}_{P}
    - \int d^{3}x\int d^{3}z \left\{A,\theta_{m}(x)\right\}_{P}
    C^{-1}_{mn}(x,z)\left\{\theta_{n}(z),B\right\}_{P} 
    \label{db}
\end{equation}
Where $C^{-1}_{mn}(x,z)$ is the inverse of the second class constraints 
PB matrix 
\[C_{mn}(x,z) \equiv \left\{\theta_{m}(x),\theta_{n}(z)\right\}.\]
In our case, $C_{mn}(x,z)$ is simply the 
$2\times 2$ bottom right corner of
(\ref{PBarray})
\begin{equation}
    C_{mn}(x,z) =
    \left( \begin{array}{c|c}
     0 & \displaystyle{-\frac{\partial\phi_{5}}
    {\partial\lambda}(x)\delta(x-z)} \\
\hline
        \displaystyle{\frac{\partial\phi_{5}}
    {\partial\lambda}(x)\delta(x-z)} &
        \displaystyle{\left[F^{i}(x)+F^{i}(z)\right]\delta_{|i}(x-z)}
    \end{array} \right) ~~ m,n = 1,2
    \label{Cmn}
\end{equation}
When dealing with field theory, the matrix $C_{mn}$ is generally a
differential operator, and the inverse matrix is not unique
unless one specifies the boundary conditions.
We choose 'no-boundary' as our boundary condition, therefore,
integration by parts can be
done freely, and the inverse matrix is
\begin{equation}
    C^{-1}_{mn}(x,z) =
    \left( \begin{array}{c|c}
    \displaystyle{\left((\frac{\partial\phi_{5}}
    {\partial\lambda})^{-2}F^{i}(x)
    + (\frac{\partial\phi_{5}}{\partial\lambda})^{-2}
    F^{i}(z)\right)\delta_{|i}(x-z)}
    & \displaystyle{\left(\frac{\partial\phi_{5}}
    {\partial\lambda}\right)^{-1}(x)\delta(x-z)} \\
\hline
        \displaystyle{-\left(\frac{\partial\phi_{5}}
    {\partial\lambda}\right)^{-1}(x)\delta(x-z)} & 0
    \end{array} \right)
    \label{C-1}
\end{equation}
The resulting DB are 
\begin{eqnarray}
    \left\{A,B\right\}_{D} & = & \left\{A,B\right\}_{P}
    + \int d^{3}x\left(\frac{\partial\phi_{5}}
    {\partial\lambda}\right)^{-2}F^{i}(x)\left[\frac{\delta A}
    {\delta\lambda(x)}(\frac{\delta B}{\delta\lambda(x)})_{|i}
    - (\frac{\delta A}{\delta\lambda(x)})_{|i}
    \frac{\delta B}{\delta\lambda(x)} \right] \nonumber \\
    & & - \int d^{3}x\left(\frac{\partial\phi_{5}}
    {\partial\lambda}\right)^{-1}(x)\left[\frac{\delta A}
    {\delta\lambda(x)}\left\{\phi_{5}(x),B\right\}
    +\left\{A,\phi_{5}(x)\right\}\frac{\delta B}{\delta\lambda(x)}
     \right]
    \label{DB}
\end{eqnarray}
This way, from now on, one should work with DB instead of PB and
take the second class constraints to vanish strongly. This will
omit the parts proportional to $P_{\lambda}$ from the first class
constraints (\ref{varphi0},\ref{varphik}) and recover 
the original form (\ref{phi0},\ref{phik}).

The algebra of the first class constraints takes the
familiar form \cite{Dirac} of a relativistic theory
\begin{mathletters}
\begin{eqnarray}
    \{\phi_{0}(x),\phi_{0}(z)\}_{D} & = &
    [h^{ij}\phi_{i}(x)+h^{ij}\phi_{i}(z)]\delta_{|j}(x-z)   \\
    \{\phi_{0}(x),\phi_{k}(z)\}_{D} & = &
    \phi_{0}(z)\delta_{|k}(x-z)  \\
    \{\phi_{k}(x),\phi_{l}(z)\}_{D} & = &
    \phi_{l}(x)\delta_{|k}(x-z) + \phi_{k}(z)\delta_{|l}(x-z)
\end{eqnarray}
\label{fDB}
\end{mathletters}
The final first class Hamiltonian of a bubble universe is
\begin{equation}
    \fbox{$\displaystyle{
    H = \int d^{3}x\left\{N^{k}y_{|k}\cdot P
    -N\frac{8\pi G}{2\sqrt{h}}\left[
    \left(\frac{\sqrt{h}}{8\pi G}\right)^{2}(\lambda + {\cal R}^{(3)})
    + P\Theta(\Psi - \lambda I)^{-1}\Theta P 
	\right]\right\} }$}
    \label{Hamiltonian}
\end{equation}
At this stage, we have a first class Hamiltonian composed of four
first class constraints, and accompanied with two second 
class constraints. The algebra of the first class constraints
is the familiar algebra of other relativistic theories.
Before moving on to quantization schemes we would like to 
study two more classical aspects: 
what happens if the action includes brane matter fields,
and what is the relation between Einstein's solutions to
the geodetic brane solutions. 


\section{Inclusion of Matter}
\label{matter}
The inclusion of matter is done by adding the action
of the matter fields to the gravitational action
\begin{equation}
    S = \int d^{4}x\left[\sqrt{-g}
    \frac{1}{16\pi G}{\cal R}^{(4)}+{\cal L}_{m}\right] ~.
\end{equation}
The matter Lagrangian density depends in general on
some matter fields, but also on the $4$-dimensional metric
$g_{\mu\nu}$. The dynamics of the matter fields
is actually not affected by the exchange of the canonical
fields from $g_{\mu\nu}$ to $y^{A}$, and one expects the
same equations of motion or the same 'matter'
Hamiltonian density.
On the other hand the momenta $P_{A}$ gets a contribution
from the matter Lagrangian
\begin{equation}
    \Delta P_{A}=\frac{\delta{\cal L}_{matter}}
    {\delta \dot{y}^{A}}=\sqrt{h}\left[ T_{nn}n^{A}
    - h^{ij}T_{ni}y^{A}_{\,,i}
    \right] ~.
\end{equation}
This contribution depends on the various projections of
the energy-momentum tensor
\begin{equation}
    T^{\mu\nu}\equiv\frac{2}{\sqrt{-g}}
    \frac{\delta{\cal L}_{matter}}{\delta g_{\mu\nu}} ~.
\end{equation}
$T_{nn}$ is the matter energy density, or the projection
of the energy-momentum
tensor twice onto the $n^{A}$ direction 
$    T_{nn} \equiv \left(T^{\mu\nu}y^{A}_{\,,\mu}y^{B}_
    {\,,\nu}\right)n_{A}n_{B} $.
While in $T_{ni}$ the energy-momentum tensor 
is projected once onto the $n^{A}$
direction and once onto the $V_{3}$ tangent space.
$    T_{ni} \equiv \left(T^{\mu\nu}y^{A}_{\,,\mu}
    y^{B}_{\,,\nu}\right)n_{A}y_{B,i}$.
See appendix \ref{app MH} for some examples of matter Lagrangians,
Hamiltonians and the
corresponding energy-momentum tensor projections.

The momenta $P_{A}$ (\ref{P2}) is now changed to
\begin{equation}
	P^{A} = -\frac{\sqrt{h}}{16\pi G}\left\{\left[
    {\cal R}^{(3)} + n_{B}\Psi^{BC}n_{C} -16\pi G T_{nn} \right]
    n^{A} + 2\Psi^{A}_{\,B}n^{B} 
+16\pi G T_{ni}h^{ij}y^{A}_{\,,i}\right\}
	\label{PT}
\end{equation}
Following the same logic that lead us from Eq.(\ref{P2}) to the
introduction of $\lambda$ (\ref{lambda}), 
we will define $\lambda$ as
\begin{equation}
        n_{A}\Psi^{A}_{\,B}n^{B} + {\cal R}^{(3)}
        -16\pi G T_{nn} + 2\lambda = 0 ~. 
	\label{lambdaT}       
\end{equation}
The effects of matter are thus 
$\lambda\rightarrow\lambda+8\pi G T_{nn}$,
$P^{A}\rightarrow P^{A}-\sqrt{h} T_{ni}h^{ij}y^{A}_{\,,i}$ but
$\Theta P$ is unchanged.
The constraints are modified as follows
\begin{eqnarray}
    \phi_{0} & \longrightarrow & \phi_{0} - \sqrt{h}T_{nn} \\
    \phi_{k} & \longrightarrow & \phi_{k} + \sqrt{h}T_{nk}  
\label{tphik}
\end{eqnarray}
Thus the Hamiltonian is changed to
\begin{equation}
    H_{G} \longrightarrow H_{G} + \int d^{3}x\sqrt{h}
    [N^{k}T_{nk}+NT_{nn}] = H_{G} + H_{m}  ~,
    \label{tH}
\end{equation}
Where $H_{m}$ is the matter Hamiltonian,
calculated in terms of the
matter fields alone as shown in appendix \ref{app MH}.
The algebra of the constraints (\ref{fDB})
remains unchanged under the inclusion
of matter, where the PB now include the derivatives
with respect to matter fields as well.


\section{The Einstein Limit}
\label{einstein}
In some manner Regge-Teitelboim gravity is a generalization of
Einstein gravity. Any solution to Einstein equations is also a solution
to RT equations (\ref{RTeq}). We will derive here the necessary
conditions for a RT-solution to be an Einstein-solution.
\begin{itemize}
\item First, we use a purely geometric relation
\begin{equation}
    2G_{nn} = {\cal R}^{(3)} + n^{B}\Psi_{BC}n^{C} ~,
    \label{Gnn}
\end{equation}
where $G_{nn}$ is the Einstein tensor twice projected onto the
$n^{A}$-direction. The constraint associated with the introduction
of $\lambda$ (\ref{lambdaT}) is
\begin{equation}
    -2\lambda = {\cal R}^{(3)} + n^{B}\Psi_{BC}n^{C} -16\pi G T_{nn}
    = 2(G_{nn}-8\pi G T_{nn})~.
\end{equation}
The Einstein solution of the equation is therefore associated with
\begin{equation}
    \fbox{$\displaystyle{\lambda = 0}$} ~.
\end{equation}
As was shown in Eq.(\ref{Psiy}), $\Psi$ has a degenerate vanishing
eigenvalue. Therefore Einstein case with $\lambda=0$, will not allow
for the essential $(\Psi-\lambda I)^{-1}$. One can not impose
$\lambda=0$ as an additional constraint (as was proposed by RT \cite{RT}),
but only look at it as a limiting case. 
\item Second, we use the projection of the Einstein
tensor once onto the $n^{A}$-direction and once onto the $V_{3}$
tangent space $G_{ni}$
\begin{equation}
    G_{ni}h^{ij}y^{A}_{\,,j} = -\Psi^{A}_{\,B}n^{B}
    -\left(y^{A}_{\,,j}(Kh^{ij}-K^{ij})\right)_{|i} ~,
    \label{Gni}
\end{equation}
in eq.(\ref{PT}) and put the momentum $P^{A}$ in the form
\begin{equation}
    P^{A} = -\frac{\sqrt{h}}{8\pi G}\left[
    (G_{nn}-8\pi GT_{nn})n^{A} - (G_{ni}-8\pi GT_{ni})
    h^{ij}y^{A}_{\,,j} + \left(y^{A}_{\,,j}
    (Kh^{ij}-K^{ij})\right)_{|i}\right] ~.
    \label{P5}
\end{equation}
It is clear that if Einstein equations $G_{nn}=8\pi GT_{nn}$ and
$G_{ni}=8\pi GT_{ni}$ are both satisfied, the momentum $P_{A}$
makes a total derivative such that
\begin{equation}
    \oint d^{3}xP_{A} = 0 ~.
    \label{Ein2}
\end{equation}
The total momentum $\oint d^{3}xP_{A}$ is a conserved
Noether charge
since the original Lagrangian does not depend
explicitly on $y^{A}$
\begin{equation}
    \mu^{A} \equiv \oint d^{3}xP^{A} = $  const.$~.
    \label{mu}
\end{equation}
The universe, as an extended object, is characterized by
the total momentum $\mu^{A}$. 
The necessary condition for an Einstein-solution is a
vanishing $\mu^{A}$.
\begin{equation}
	  \fbox{$\displaystyle{
    \mu^{A} \equiv \oint d^{3}xP^{A} = 0 }$}~.
	\label{mu0}
\end{equation}
\item The condition (\ref{mu0}) simply tells us 
that the total 'bulk' momentum of the
universe vanishes. This motivates us to use a new coordinate
system for the embedding, namely, the 'center of mass frame' $+$
'relative coordinates'. As relative coordinates we will use the 
derivatives $y^{A}_{\,,i}\,$. This has a direct relation to the metric
and therefore, we expect the equation of motion to resemble Einstein's equations. 
The new system and the calculations appear in appendix \ref{app DM}.
\end{itemize}


\section{Quantization}
\label{quantization}
The treatment so far was classical, but the derivation of the 
Hamiltonian and the construction of the various constraints
are the ingredients one needs for quantization. 
In the following sections we will describe two quantization 
schemes, canonical quantization and functional integral
quantization. 
\subsection{Canonical Quantization}
Dirac's procedure leads us towards the canonical quantization of
our constrained system. The following recipe was constructed by
Dirac \cite{Dirac} for quantizing a constrained system within the
Schrodinger picture
\begin{itemize}
    \item Represent the system with a state vector
        (wave functional) .
    \item Replace all observables with operators.
    \item Replace DB with commutators,
    $\displaystyle{\left\{,\right\}_{D} \longrightarrow i\hbar
    \left[,\right]}$
    \item First class constraints annihilate the state vector.
    \item Second class constraints represent operator identities.
    \item Since the commutator is ill defined for fields at the 
	same space point, one must place all momenta to the right
	of the constraint. 
    \item First class constraints must commute with each other.
        This ensures consistency, and may call for operator
        ordering within the constraint.
\end{itemize}
In our case, we can use the coordinate representation. The state
vector is represented by a wave functional $\Phi[y]$. The
DB (commutator) between $y^{A}$ and $P_{B}$ is canonical,
therefore, these operators can be represented in a canonical way
\begin{eqnarray}
    & & \hat{y}^{A}(x) \Rightarrow y^{A}(x) \nonumber \\
    & & \hat{P}_{A}(x) \Rightarrow -i\hbar\frac{\delta}
    {\delta y^{A}(x)} \nonumber
\end{eqnarray}
The operator $\hat{P_{\lambda}}$ vanishes identically.
The DB of $\lambda$ with $y^{A}$,$P_{B}$ are not canonical,
therefore the operator $\hat{\lambda}$ must be expressed as a function
of $\hat{y^{A}}$,$\hat{P_{B}}$ . This can be done with the aid of the second
class constraint (\ref{theta2}).

The first class constraints as operators must annihilate the wave
functional. These constraints are recognized as 
\begin{enumerate}
\item The momentum constraint (\ref{varphik})
\begin{equation}
    -i\hbar y^{A}_{\,|k}\frac{\delta\Phi}
    {\delta y^{A}} = 0 ~,
    \label{momcons}
\end{equation}
which simply means that the wave functional is a $V_{3}$ scalar
and does not change its value under reparametrization of the
space coordinates. This can be shown if one takes an infinitesimal
coordinate transformation
\begin{eqnarray}
    x^{k} & \longrightarrow & x^{k}+\epsilon^{k} ~,\nonumber \\
    y^{A}(x) & \longrightarrow & y^{A}(x)+
    \epsilon^{k}y^{A}_{\,|k}(x) ~,\nonumber \\
    \Phi[y] & \longrightarrow & \Phi[y] +
    \epsilon^{k}y^{A}_{\,|k}\frac{\delta\Phi[y]}
    {\delta y^{A}} ~.\nonumber
\end{eqnarray}
The wave functional is unchanged if and only if the momentum
constraint holds.

\item The other constraint is the Hamiltonian constraint, and up to
order ambiguities, the equation
is the analog to the Wheeler de-Witt equation
\begin{equation}
    \frac{8\pi G}{2\sqrt{h}}\left[
    \left(\frac{\sqrt{h}}{8\pi G}\right)^{2}
        (\hat{\lambda} + {\cal R}^{(3)})(x)
        -\hbar^{2}\left((\Psi - \hat{\lambda}I)^{-1}\right)^{AB}(x)
        \frac{\delta^{2}}
    {\delta y^{A}(x)\delta y^{B}(x)}\right]\Phi[y] = 0 ~.
    \label{WdW}
\end{equation}
It is accompanied however, with the operator identity 
\begin{equation}
	\frac{8\pi G}{2\sqrt{h}}\left[
    \left(\frac{\sqrt{h}}{8\pi G}\right)^{2}
    + \hat{P}\Theta(\Psi - \hat{\lambda} I)^{-2}\Theta 
\hat{P}\right] =0
\end{equation}
\end{enumerate}


\subsection{Functional Integral Quantization}
Calculating functional integrals for a constrained 
system is not new. This was done for first class 
constraints by BFV \cite{BFV}, And was
generalized for second class constraints by 
Fradkin and Fradkina \cite{FF}.

The first step is actually a classical calculation, 
that is, calculating the BRST generator \cite{BRST}. 
For this calculation we will adopt the 
following notations:
\begin{itemize}
	\item The set of canonical fields will include 
	the Lagrange multipliers $N^{\mu}=(N,N^{i})$ that is
	$Q^{A}=\left( y^{A}, \lambda, N^{\mu}\right)^{T}$, 
and the corresponding 
	conjugate momenta $\Pi_{A}=( P_{A}, P_{\lambda}, \pi_{\mu})$.
	The Lagrange multipliers are not dynamical, therefore, the 
	conjugate momenta must vanish. This doubles the number of 
	first class constraints $G_{a}=(\pi_{\mu}, \phi_{\nu})$. 
	\item For each constraint we introduce a pair of fermionic 
	fields $\eta^{a}=\left( \rho^{\mu}, c^{\mu}\right)^{T}$, 
and the conjugate momenta 
	${\cal P}_{a}=(\bar{c}_{\nu},\bar{\rho}_{\nu})$. 
	(In our case, all constraints are bosonic, therefore
	the ghost fields are fermions).
	\item Each index actually represent a discrete index and a 
	continuous index, for example, $y^{A}\equiv y^{A}(x)$. The
	summation convention is then generalized to sum over the
	continuous index as well
	\begin{equation} 
	N^{\mu}\phi_{\mu}\equiv \int d^{3}x N^{\mu}(x)\phi_{\mu}(x)~.
	\end{equation}
	\item We use Dirac Brackets as in (\ref{DB}), but 
	the Poisson Brackets are generalized to include bosonic
	and fermionic degrees of freedom
	\begin{equation}
    	\left\{L,R\right\} =
    	\frac{\partial^{r} L}{\partial q^{A}}
	\frac{\partial^{l} R}{\partial p_{A}}
	-(-1)^{n_{L}n_{R}}\frac{\partial^{r} R}{\partial q^{A}}
	\frac{\partial^{l} L}{\partial p_{A}} ~.
	\end{equation}
 	Where $(q,p)$ is the set of canonical fields including 
	the fermionic fields. $r,l$ denote right and left 
	derivatives
	\begin{equation}
    	dR=\frac{\partial^{r} R}{\partial q}dq
	=dq\frac{\partial^{l} R}{\partial q}~.
	\end{equation}
	And the fermionic index is
  	\begin{equation}
    	n_{R}=\left\{\begin{array}{l l} 0 & $if R is a boson$ \\
			1 & $if R is a fermion$ \end{array} \right.  
	\end{equation}
\end{itemize}
Let us now calculate the structure functions of the theory.
The first order structure functions are defined by the algebra
of the constraints $\{G_{a},G_{b}\}_{D}=G_{c}U^{c}_{ab}$.
It is only the original constraints, 
(not the multipliers momenta), that have non vanishing 
structure functions (\ref{fDB}). 
\begin{equation}
	\left\{\left(\begin{array}{c}\pi_{\mu}(x)\\ 
	\phi_{\mu}(x) 
	\end{array}\right)
	,\left(\pi_{\nu}(z), \phi_{\nu}(z)\right)\right\}_{D}  = 
	\left(\begin{array}{c c} 0 & 0 \\ 0 & 
	\int d^{3}w\,\phi_{\lambda}(w)
	U^{\lambda}_{\mu\nu}(x,z,w) \end{array}\right) ,
\end{equation}
and the relevant first order structure functions are
\begin{equation}
	U^{\lambda}_{\mu\nu}(x,z,w) =  
	\left[\delta^{0}_{\mu}\delta^{0}_{\nu}h^{\lambda k}
	\left(\delta(w-x)+\delta(w-z)\right)
	+\delta^{\lambda}_{\mu}\delta^{k}_{\nu}\delta(w-z)
	+\delta^{k}_{\mu}\delta^{\lambda}_{\nu}\delta(w-x) 
	\right]\delta_{,k}(x-z) 
	\label{fosf}
\end{equation}
(Generally, one should also look at 
$\{H_{0},G_{a}\}_{D}=G_{b}V^{b}_{a}$, but here $H_{0}=0$).
The second order structure functions are defined by the 
Jacobi identity $\protect{{\cal A}(\{\{G_{a},G_{b}\}_{D},G_{c}\}_{D})=0}$,
where ${\cal A}$ means antisymmetrization. Using the first 
order functions (\ref{fosf}) one gets
$\protect{{\cal A}(G_{d}[\{U^{d}_{ab},G_{c}\}_{D}+ U^{d}_{ec}U^{e}_{ab}])=0}$.
This equation is satisfied if and only if the expression in 
the square brackets is again a sum of constraints
\begin{equation}
	{\cal A}(\{U^{d}_{ab},G_{c}\}_{D}+ U^{d}_{ec}U^{e}_{ab})
	= G_{f}U^{fd}_{abc} 
	\label{sosf}.
\end{equation} 
The second order structure functions $U^{fd}_{abc}$ are 
antisymmetric on both sets of indices. 
In our case, the second order structure functions vanish,
and the theory is of rank $1$.
This resembles ordinary gravity and string theory as oppose to 
membrane theory, where the rank is the dimension of the 
underlying space manifold.
The BRST generator of a rank $1$ theory is given by
$\Omega = G_{a}\eta^{a} + \frac{1}{2}{\cal P}_{c}
	U^{c}_{ab}\eta^{b}\eta^{a}$.
Here it is 
\begin{equation}
	\Omega = \int d^{3}x[\pi_{\mu}\rho^{\mu}+ 
	\phi_{\mu}c^{\mu}+ h^{kl}\bar{\rho}_{k}
	c^{0}_{\,,l}c^{0} +\bar{\rho}_{\mu}
	c^{\mu}_{\,,k}c^{k}](x).
	\label{Omega}
\end{equation}

The main theorem of BFV \cite{BFV} is that 
the following functional integral
does not depend on the choice of the 
gauge fixing Fermi function $\Psi$ :
\begin{equation}
    Z_{\Psi} = \int {\cal D}Q^{A}{\cal D}\Pi_{A}
	{\cal D}\eta^{a}{\cal D}{\cal P}_{a}\,M\,
	\exp[i \int dt(\Pi_{A}\dot{Q}^{A} + 
	{\cal P}_{a}\dot{\eta}^{a} - H_{\Psi})]~.
	\label{Z1}
\end{equation}
Where $M=\delta(\theta_{1})\delta(\theta_{2})(\det C_{mn})^{1/2}$
is taking care of the second class constraints, and, since
the canonical Hamiltonian vanishes, 
$H_{\Psi}=-\{\Psi,\Omega\}_{D}$. 

The determinant of $C_{mn}$ for compact space manifolds,
is calculated in a simple 
way in appendix \ref{app detC}.

\section{An Example: Geodetic Brane Quantum Cosmology}
\label{mini}
In the following example we would like to implement
GBG to cosmology, and in particular to quantum cosmology.
Detailed examples and calculations can be found in 
\cite{RTcos,minipath}, here we will just focus on global 
characteristics of the Feynman propagator for a 
geodetic brane within the minisuperspace model.
Attention will be given to the differences between 'Geodetic
Brane Quantum Cosmology' and the standard 'Quantum Cosmology'. 

The standard and simple way to describe
the cosmological evolution of the universe is to assume
that on large scales the universe is homogeneous and
isotropic. The geometry of such a universe is described by
the Friedman-Robertson-Walker (FRW) metric
\begin{equation}
    ds^{2}=-N^{2}(t)dt^{2}+a^{2}(t)d\Omega_{3}^{2},
    \label{FRW}
\end{equation}
where $N(t)$ is the lapse function, $a(t)$ is the scale
factor of the universe, and
\begin{equation}
    d\Omega_{3}^{2}=d\psi^{2}+\chi^{2}(\psi)d\Omega_{2}^{2}
    \label{dOmega3}
\end{equation}
is the line element of the $3$ dimensional spacelike hypersurface
which is assumed to be homogeneous and isotropic.
$d\Omega_{2}^{2}$ is the usual line element on a $2$ sphere,
and \mbox{$\chi(\psi)=\sin\psi\,/\,\psi\,/\,\sinh\psi$}
if the $3$ space
is closed, flat or open respectively.
In General Relativity, the components of the metric are
the dynamical fields, the lapse function $N(t)$ is actually
a Lagrange multiplier, and the only dynamical variable
is the scale factor $a(t)$. 
This model is called minisuperspace, since, 
the infinite number of
degrees of freedom in the metric is reduced to a
finite number.
The remnant of general
coordinate transformation invariance, is time
reparameterization invariance, that is, the arbitrariness
in choosing $N(t)$. The usual and most convinient gauge
is $N=1$.

In GBG the situation is quite different. First, one has to embed
the FRW metric (\ref{FRW}) in a flat manifold.
The minimal embedding of a FRW metric calls for
one extra dimension. We will work
here, for simplicity, with the closed universe $\chi=\sin\psi$.
The embedding in a flat Minkowski spacetime
with the signature $(-,+,+,+,+)$, is given by \cite{rosen}
\begin{equation}
    y^{A}=\left(\begin{array}{l}T(t)
    \\a(t)z^{I}(x)\end{array}\right) \;\;\;
    z^{I}=\left(\begin{array}{l}\sin\psi\sin\theta\cos\phi\\
        \sin\psi\sin\theta\sin\phi\\
       \sin\psi\cos\theta\\ \cos\psi
        \end{array}\right)~.
    \label{5embed}
\end{equation}
The lapse function is given by $N(t)=\sqrt{\dot{T}^{2}-\dot{a}^{2}}$,
it is {\em not} a Lagrange multiplier, but it depends on the two
dynamical variables: the scale factor $a(t)$ and the external
timelike coordinate $T(t)$. Time reparameterization invariance is,
naturally, an intrinsic feature of
\mbox{$\sqrt{\dot{T}^{2}-\dot{a}^{2}}\,dt$}, but, no gauge fixing is
allowed here, since both $T(t)$ and $a(t)$ are dynamical.
The gravitational Lagrangian (\ref{lang2}), after integrating
over the spatial manifold is
\begin{equation}
    L=\sigma\left(3Na-\frac{3a\dot{a}^{2}}{N}\right).
    \label{minilang}
\end{equation}
$\sigma=\frac{2\pi^{2}}{8\pi G}$ is a scaling factor,
for convinience we will set $\sigma=1$.
The key for quantization is of course the Hamiltonian.
One can derive the Hamiltonian directly from the
Langrangian (\ref{minilang}), or, to use the ready made
Hamiltonian (\ref{Hamiltonian}) and just insert the
'minimized' expressions for the embedding vector, and
the conjugate momenta.

\subsection{Minisuperspace Hamiltonian}
The first step is to introduce the coordinates and
conjugate momenta. The general embedding vector
$y^{A}$ is replaced by the dynamical degrees
of freedom $a(t)$ and $T(t)$, while the spatial
dependence is forced by the expression (\ref{5embed}).
It is expected that the conjugate momenta will have
two degrees of freedom $P_{a}(t),P_{T}(t)$, the
delicate issue is the spatial dependence of the
momenta. Our choice is
\begin{equation}
    P_{A}=\left(\begin{array}{l}P_{T}(t)
    \\P_{a}(t)z^{I}(x)\end{array}\right)\cdot
    \frac{\sin^{2}\psi\sin\theta}{8\pi G} ~,
    \label{5momenta}
\end{equation}
the factor $\sin^{2}\psi\sin\theta$ was inserted
in order to keep the momenta
a $3$-dimensional vector density. The
spatial dependence is through $z^{I}(x)$ such that
the momentum constraint (\ref{phik}) vanishes strongly.
And, the normalization is \mbox{$\int d^{3}x\,\dot{y}^{A}p_{A}
=\sigma(\dot{a}P_{a}+\dot{T}P_{T})$}.
In addition, we set $\lambda=\lambda(t)$ and
$P_{\lambda}=P_{\lambda}(t)\frac{\sin^{2}
\psi\sin\theta}{8\pi G}$.

Inserting these expressions into the constraints
(\ref{phis}) and integrating over spatial coordinates,
one is left with one first class constraint
\begin{equation}
    \varphi=\frac{1}{2}\left(6a+a^{3}\lambda
    +\frac{P_{T}^{2}}{a^{3}\lambda}
    +\frac{P_{a}^{2}}{6a-a^{3}\lambda}
    +\alpha P_{\lambda}\right)\approx 0 ~,
    \label{mini1cons}
\end{equation}
and two second class constraints
\begin{eqnarray}
    \theta_{1}&=&P_{\lambda}\approx 0 \nonumber\\
    \theta_{2}&=&\frac{1}{2}\left(a^{3}
    -\frac{P_{T}^{2}}{a^{3}\lambda^{2}}
    +\frac{a^{3}P_{a}^{2}}{(6a-a^{3}\lambda)^{2}}\right)\approx 0~.
    \label{mini2cons}
\end{eqnarray}
The Dirac brackets (\ref{DB}) are defined as
\begin{eqnarray}
    \left\{A,B\right\}_{D} & = & \left\{A,B\right\}_{P}
    -\left(\frac{P_{T}^{2}}{a^{6}\lambda^{3}}
    +\frac{a^{6}P_{a}^{2}}{(6a-a^{3}\lambda)^{3}}\right)^{-1}
    \left[\frac{\partial A}{\partial\lambda}\{\theta_{2},B\}
    +\{A,\theta_{2}\}\frac{\partial B}{\partial\lambda}\right]
\end{eqnarray}
and the minisuperspace Hamiltonian is
\begin{equation}
    H=\frac{-N}{2}\left(6a+a^{3}\lambda
    +\frac{P_{T}^{2}}{a^{3}\lambda}
    +\frac{P_{a}^{2}}{6a-a^{3}\lambda}
    +\alpha P_{\lambda}\right) ~,
    \label{minihamiltonian}
\end{equation}
We would like to focus on the Feynman propagator \cite{Feynman}
$K(a_{f},T_{f},t_{f};a_{i},T_{i},t_{i})$
for the empty geodetic brane universe.
Although the empty universe is a
non-realistic model for our universe, the calculation
of the propagator is simple and it demonstrates some
of the main features and advantages of Geodetic Brane
Quantum Cosmology over the standard quantum cosmology models.
This propagator is the
probability amplitude that the universe is in
$(a_{f},T_{f})$ at time $t_{f}$, and it was in $(a_{i},T_{i})$
at time $t_{i}$. 
We will use a modified
version of BFV integral offered by Senjanovic \cite{senj},
where the ghosts and multipliers were integrated out.
\begin{eqnarray}
    K(a_{f},T_{f},t_{f};a_{i},T_{i},t_{i})&=&\int d\mu\,
    \exp\left[2\pi i\int_{t_{i}}^{t_{f}}dt(\dot{a}P_{a}+\dot{T}P_{T}
    +\dot{\lambda}P_{\lambda})\right]
    \nonumber \\
    d\mu&=& da\,dP_{a}\,dT\,dP_{T}\,d\lambda\,dP_{\lambda}\,
    \delta(\varphi)\,\delta(\chi)\,|\{\chi,\varphi\}|\,
    \delta(\theta_{1})\,\delta(\theta_{2})\,
    |\det(\{\theta_{m},\theta_{n}\})|^{1/2}
    \label{prop1}
\end{eqnarray}
This propagator is calculated in phase space, where
the measure is the Liuville measure $dx\,dp$.
In addition, the measure $d\mu$ enforces the constraints
(first and second class) by delta functions, it
includes an arbitrary gauge fixing function $\chi$,
the determinants of the Poisson brackets between
first class constraints and the gauge fixing function
and the determinants of the Poisson brackets 
between second class constraints.
Attention should be given to the following issues:  
\begin{itemize}
\item The canonical Hamiltonian vanishes, therefore
    it is absent in the action.
\item The boundary conditions for the propagator
    determine the values of $a_{f},T_{f},a_{i},T_{i}$,
    but not the value of $\lambda$ nor the values of
    the momenta. Therefore, the momenta and $\lambda$
    must be integrated over at the initial point.
\item The gauge fixing function $\chi$, although arbitrary,
    must be chosen such that it does not violate the
    boundary conditions nor the constraints.
    In addition, the Poisson brackets $\{\chi,\varphi\}$
    must not vanish.
\item The determinant of the second class constraints 
Poisson brackets is simply
\begin{equation}
    |\det(\{\theta_{m},\theta_{n}\})|^{1/2}
    =\left|\frac{\partial\theta_{2}}{\partial\lambda}\right|
    =\left|\frac{P_{T}^{2}}{a^{3}\lambda^{3}}
    +\frac{a^{6}P_{a}^{2}}{(6a-a^{3}\lambda)^{3}}\right|
    \label{theta2,lambda}
\end{equation}
\item Our convention here is $\sigma=1$ and Planck constant
    $h=1$ ($\hbar=\frac{1}{2\pi}$).
\item In cases where matter is included, 
the inclusion of matter will affect in a few places.
The action will include terms like $\dot{\phi}\pi$,
an integration over matter fields and momenta will be added,
and the first class constraint will have a contribution
which is simply the matter Hamiltonian
\mbox{$\varphi\longrightarrow\varphi+H_{m}(a,\phi,\pi)$.}
All other constraints remain intact.
\end{itemize}
The calculation of the propagator (\ref{prop1}) is
carried out in a simple way following Halliwell \cite{halliwell},
and the final propagator takes the form
\begin{equation}
    K_{\pm}(a_{f},T_{f};a_{i},T_{i})
    =\int d\omega\,
    \exp\left[2\pi i\omega(T_{f}-T_{i})
    \mp2\pi i\omega^{2}\left(F(x_{f})-F(x_{i})\right)\right]
    \label{prop3}
\end{equation}
The index of $K_{\pm}$ and the $\mp$ in the exponent
refers to the expanding/contracting scale factor.
$\omega$ is the conserved bulk energy 
(the momentum conjugate to the bulk time coordinate $T$).
Since the value of $\omega$ is not fixed at the initial condition,
one must integrate over $\omega$. 
One should notice according to eq.(\ref{mu0}) 
that the Einstein solution is assiciated with $\omega=0$.
The function $F(x)$ is given by
\begin{equation}
    F(x)=\left\{\begin{array}{ll}
    \frac{1}{12}\left[3\text{Arcsin} x
    +\sqrt{1-x^{2}}(4x^{5}+2x^{3}-3x)\right] &
    |x|\leq 1 \\
    \text{sgn}(x)\frac{\pi}{8}-\frac{i}{12}\left[3\text{sgn}(x)
    \text{Arccosh}|x|-\sqrt{x^{2}-1}(4x^{5}+2x^{3}-3x)\right]
    & 1<|x| \end{array}\right.
    \label{F}
\end{equation}
Where $x=(\frac{3a}{\omega})^{1/3}$.

Let us now examine the properties of the propagator (\ref{prop3}).
Actually, the propagator is independent of the internal time
parameter $t$ (a common character of all parameterized theories), 
and depends exclusively on the value
of $a$ and $T$ at the boundaries. 
\begin{itemize}
\item
The most basic characteristic of a propagator is
the possibility to propagate from an initial state
to a final state through an intermidiate state. For
example, the propagator for a non-relativistic particle is
\mbox{$K(x_{3},t_{3};x_{1},t_{1})=\int dx_{2}
K(x_{3},t_{3};x_{2},t_{2})K(x_{2},t_{2};x_{1},t_{1})$.}
At the intemidiate time $t_{2}$, one must integrate over
$x_{2}$. It is clear that there is no integration over $t_{2}$,
$t$ is the evolution parameter, it must be monotonic
$t_{3}>t_{2}>t_{1}$, and integration over $t_{2}$ makes no sense.
Another characteristic of the propagator is
\mbox{$\lim_{t_{2}\rightarrow t_{1}}K(x_{2},t_{2};x_{1},t_{1})
=\delta(x_{2}-x_{1})$}. 
The situation with parameterized theories is quite different.
The propagator is independent of the internal time, 
and integration over all dynamical variable diverges.
The solution is, usually, to let one of the dynamical 
variables as 'time', and integrate only over the other variables.

The question is: How does the propagator (\ref{prop3})
behaves at the intermidiate point ? What is the relevant
evolution parameter and what integrations should be made ?
One can check that if $a$ is taken to be the monotonic
evolution parameter and integration over $T$ at the intermidiate
point is done, then the propagator (\ref{prop3}) is well
behaved.
\begin{eqnarray}
    K(a_{3},T_{3};a_{1},T_{1})&=&
    \int dT_{2}\int d\omega\,e^{2\pi i[\omega(T_{3}-T_{2})
    -\omega^{2}(F(x_{3})-F(x_{2}))]}\int d\bar{\omega}\,
    e^{2\pi i[\bar{\omega}(T_{2}-T_{1})
    -\bar{\omega}^{2}(F(\bar{x}_{2})-F(\bar{x}_{1}))]} \nonumber \\
    &=& \int d\omega\,e^{2\pi i[\omega(T_{3}-T_{1})
    -\omega^{2}(F(x_{3})-F(x_{1}))]} \\
    K(a_{1},T_{2};a_{1},T_{1})&=&
    \int d\omega\,e^{2\pi i\omega(T_{2}-T_{1})}=\delta(T_{2}-T_{1})
    \label{K32K21}
\end{eqnarray}
This cannot be done within the standard quantum cosmology
models, since there, the only dynamical variable is the scale
factor $a$. Such a propagator of only one variable contains 
no information, it can tell that the varible is monotonic.
The common solution in standard quantum cosmology
is to add another dynamical variable such
as a scalar field and to use one of them as the evolution
parameter. Here we see one of the main advantages of 
Geodetic Brane Quantum Cosmology over the standard models, 
the problem of time has an intrinsic solution as we have one
extra degree of freedom which serves as 'time'.
\item
The most general wave function that can be generated using
the propagator (\ref{prop3}) is
\begin{equation}
    \Psi(a,T)=\int d\omega\,e^{2\pi i\omega T}\left[
    A(\omega)e^{-2\pi i \omega^{2}F(x)}
    +B(\omega)e^{2\pi i \omega^{2}F(x)}\right]~.
    \label{wavefunction}
\end{equation}
One can verify that the wave function (\ref{wavefunction})
(and the propagator (\ref{prop3})) satisfy the
corresponding WDW equation
\begin{equation}
    \hbar^{2}\left[-\xi(x)\frac{\partial}{\partial a}
    \left(\frac{1}{\xi(x)}\frac{\partial}{\partial a}\right)
    +\xi^{2}(x)\frac{\partial^{2}}{\partial T^{2}}\right]\Psi(a,T)=0
    \label{miniWDW}
\end{equation}
Where $\xi(x)=(1+2x^{2})\sqrt{1-x^{2}}$, and
\mbox{$x=\left(3a(-i\hbar\frac{\partial}{\partial T})^{-1}\right)^{1/3}$.}
Putting $-i\hbar\frac{\partial}{\partial T}=\omega$ and neglecting
the term proportional to the first derivative
$\frac{\partial \Psi}{\partial a}$, the equation
(\ref{miniWDW}) looks like a zero energy Schrodinger equation
\begin{equation}
    \left[-\hbar^{2}\frac{\partial^{2}}{\partial a^{2}}
    +V_{\omega}(a)\right]\Psi_{\omega}(a)=0~,
    \label{minischr}
\end{equation}
with the potential
\begin{equation}
    V_{\omega}(a)=-\omega^{2}\left[1-\left(\frac{3a}
	{\omega}\right)^{2/3}\right]
    \left[1+2\left(\frac{3a}{\omega}\right)^{2/3}\right]^{2}
    =36a^{2}-3\omega^{4/3}(3a)^{2/3}-\omega^{2}
    \label{minipotential}
\end{equation}
\begin{figure}[tbp]
    \begin{center}
        \includegraphics[scale=0.7]{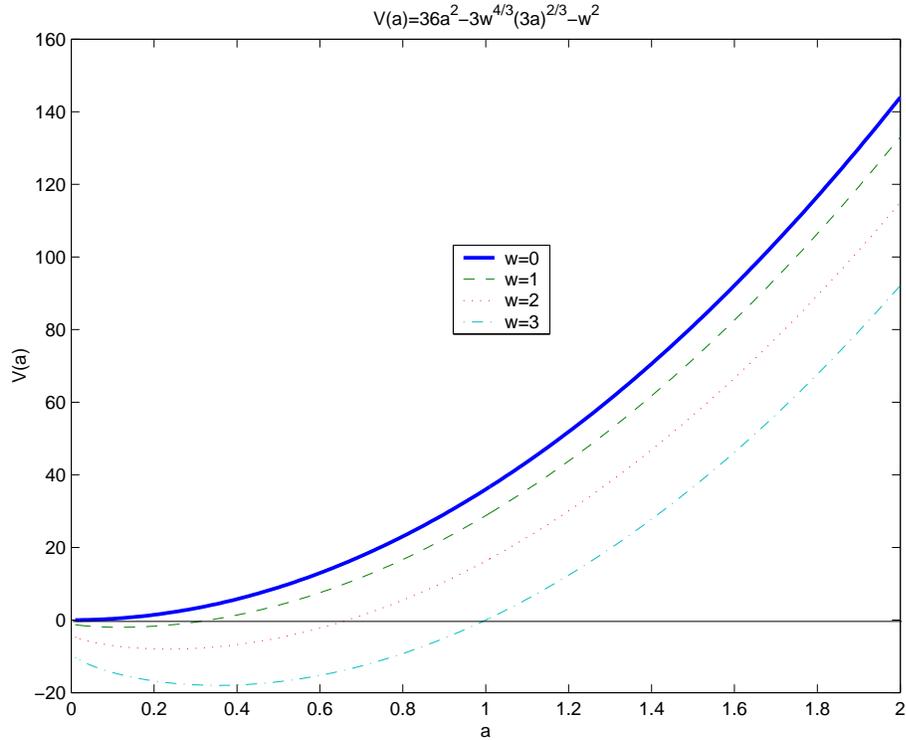}
        \caption['Schrodinger' potential for a minisuperspace model]
        {The potential $V_{\omega}(a)$}
        \label{fig.potential}
    \end{center}
\end{figure}
The classical turning point is $a=\omega/3$, and the empty
brane universe can not expand classically byeond this point.
The empty universe model is non-realistic, a more realistic
model may include some matter fields, or at least
a cosmological constant. Analysis of the cosmological
constant universe can be found in \cite{RTcos}.
\item
In accordance with section \ref{einstein}, one of the
necessary conditions for an Einstein solution is Eq.\
(\ref{Ein2}) $\int d^{3}x\,P_{A}=0$. Within our
minisuperspace model, integrating the momenta (\ref{5momenta})
over the spatial manifold one gets
$\int d^{3}x\,P_{A}=(P_{T},0,0,0,0)^{T}$,
thus the Einstein case is associated with $\omega=0$,
and the only classcal regime is $a=0$.
\item
The still open question is that of the boundary conditions.
In particular $\Psi(a=0,T)$ and $\Psi(a\rightarrow\infty,T)$.
One possibility is: $\Psi$ vanishes at the big bang ($a=0$)
and $\Psi$ is bounded at $a\rightarrow\infty$. This will lead
to $\omega$ quantization $\omega_{n}^{2}=8\hbar(n+1/4)$
where $n$ is a positive integer.
Clearly, the Einstein case $\omega=0$ is excluded by such
quatization condition.
\end{itemize}

\section*{summary}
\begin{enumerate}
\item In the present model of Geodetic Brane Gravity, 
the $4$ dimensional universe floats as an extended object
within a flat $m$ dimensional manifold. It can be 
generalized however, to include fields in the 
surrounding manifold (bulk), this is done by adding
the bulk action integral to the action of the brane.
The brane will feel those bulk fields as forces 
influencing its motion \cite{carter}. 
The bulk fields may include matter fields or the bulk
gravity \cite{RS,Dvali,Ida,Stealth}. 
\item In this paper we have derived the quadratic 
Hamiltonian of a brane universe.
The Hamiltonian is a sum of $4$ first class constraints, 
while $2$ additional
second class constraints are present. 
We used Dirac Brackets and found the 
algebra of first class constraints to 
be the familiar one from other 
relativistic theories (such as string, 
membrane or general relativity).
The BRST generator turns out to be of rank $1$.
\item Geodetic brane gravity modifies 
general relativity, and introduces
in a natural way {\em dark matter} components.
Dark matter in inflationary models which accompanies
ordinary matter to govern the evolution of the universe
can be found in \cite{DKL}. 
\item We have formulated the conditions
for a solution to be that of general 
relativity, and showed that the 
Einstein case can be achieved only as a limiting case.
\item Canonical quantization is possible 
with the aid of Dirac brackets.
The resulting Wheeler de-Witt equation 
includes operators which are
not free, but are constrained by 
the second class constraints 
as operator identities.
\item The ground is ready for 
functional integral quantization, 
the BRST generator is of rank $1$, 
and the determinant of second 
class constraints has been brought to a simple form. 
\item A simple application of geodetic brane gravity 
to cosmology is possible within 
the framework of a minisuperspace.
Classical cosmological models appear in \cite{davidson,higgs}.
Canonical quantization appears in \cite{RTcos}, and the complementary
functional integral quantization in \cite{minipath}.
\item Another significant advantage of GBG over GR 
is the solution to the problem of time. While a homogeneous and 
isotropic metric is characterized by only one dynamical 
variable (the scale factor of the universe), the embedding
vector contains two dynamical variables (the scale factor and 
the bulk time). Thus, taking the embedding vector to be the 
canonical variables, will enhance the theory with one extra
variable that may be intepreted as a time coordinate.    
\end{enumerate}


\appendix

\section{Functional Derivatives}
\label{app FD}
\noindent $\bullet$ Let $F[y]$ be a functional of $y(x)$ such that
$\delta F = \int d^{3}x f(x)\delta y(x)$  then the functional
derivative is $\displaystyle{\frac{\delta F}{\delta y(x)} \equiv f(x)}$.

The chain rule holds for functional derivatives
$\displaystyle{\frac{\delta F(G[y])}{\delta y(x)}
=\frac{\partial F}{\partial G}\frac{\delta G[y]}{\delta y(x)}}$

\noindent $\bullet$ The delta distribution is a scalar density
of weight $1$ such that for a $3$-scalar $f(x)$
\begin{equation}
    f(x)=\int d^{3}z f(z)\delta^{3}(x-z)
\end{equation}
The covariant derivative of the delta function
$\delta^{3}_{\,|i}(x-z)$ is defined for a $3$-vector $g^{i}(x)$
\begin{equation}
    \int d^{3}x g^{i}(x)\delta^{3}_{\,|i}(x-z) = -g^{i}_{\,|i}(z).
\end{equation}

\noindent $\bullet$ The delta function is symmetric with its two
arguments
\begin{equation}
    \delta(x-z) = \delta(z-x) ~.
\end{equation}
The first covariant derivative of the delta function is
antisymmetric with its arguments
\begin{equation}
        \delta_{\,|i}(x-z)\equiv \nabla_{x^{i}}\delta(x-z) =
    -\nabla_{z^{i}}\delta(z-x) \equiv -\delta_{\,|i}(z-x) ~.
\end{equation}
While the second covariant derivative is again symmetric.

\noindent $\bullet$ The basic functional derivatives are
\begin{eqnarray}
    \frac{\delta y^{A}(x)}{\delta y^{B}(z)} &=&
    \delta^{A}_{B}\delta(x-z) \\
    \frac{\delta y^{A}_{\,|i}(x)}{\delta y^{B}(z)} &=&
    \delta^{A}_{B}\delta_{|i}(x-z) \\
    \frac{\delta y^{A}_{\,|ij}(x)}{\delta y^{B}(z)} &=&
    \left(\delta^{A}_{B}-y^{A}_{\,|a}h^{ab}y_{B|b}\right)
    \delta_{|ij}(x-z) -
    y_{B|ij}y^{A}_{\,|a}h^{ak}\delta_{|k}(x-z)
\end{eqnarray}
For a general expression $\Phi(x,y,y_{|i},y_{|ij})$
the functional derivative is
\begin{eqnarray}
    \frac{\delta \Phi(x)}{\delta y^{A}(z)} & = &
    \frac{\partial \Phi}{\partial y^{A}}(x)\delta(x-z)
    + \frac{\partial \Phi}{\partial y^{A}_{\,|i}}(x)
    \delta_{|i}(x-z)  \nonumber \\
    & + & \frac{\partial \Phi}{\partial y^{B}_{\,|ij}}(x)
    \left[\left(\delta^{B}_{A}-y^{B}_{\,|b}h^{ab}y_{A|a}\right)
    \delta_{|ij}(x-z) -
    y_{A|ij}y^{B}_{\,|b}h^{bk}\delta_{|k}(x-z)\right]
\end{eqnarray}
Another nontrivial example is the $3$-dimensional Christofel symbols
$\Gamma^{i}_{kl}=h^{ij}y^{A}_{\,,j}y_{A,kl}$
\begin{equation}
        \frac{\delta \Gamma^{i}_{kl}(x)}{\delta y^{A}(z)} =
    h^{ij}y_{A|kl}(x)\delta_{|j}(x-z)
    + h^{ij}y_{A|j}(x)\delta_{|kl}(x-z) ~.
\end{equation}

\noindent $\bullet$ The Poisson brackets are defined 
in the usual way
\begin{equation}
    \left\{F,G\right\} =
    \int d^{3}x \left(\frac{\delta F}{\delta y^{A}(x)}
    \frac{\delta G}{\delta P_{A}(x)} - \frac{\delta F}{\delta P_{A}(x)}
    \frac{\delta G}{\delta y^{A}(x)} \right) ~.
    \label{PB}
\end{equation}

\section{Poisson Brackets of constraints}
\label{app DB}
We will start with the constraints (\ref{phis})
\begin{mathletters}
\begin{eqnarray}
    \phi_{0} & = & \frac{8\pi G}{2\sqrt{h}}\left[
    \left(\frac{\sqrt{h}}{8\pi G}\right)^{2}(\lambda + {\cal R}^{(3)})
    + P\Theta(\Psi - \lambda I)^{-1}\Theta P 
    \right]\approx 0 ~, \\
    \phi_{k} & = & y_{\,|k}\cdot P \approx 0 ~, \\
    \phi_{4} & = & P_{\lambda} \approx 0 ~, \\
    \phi_{5} & = & \frac{8\pi G}{2\sqrt{h}}\left[
    \left(\frac{\sqrt{h}}{8\pi G}\right)^{2}
    + P\Theta(\Psi - \lambda I)^{-2}\Theta P 
    \right]\approx 0  ~.
\end{eqnarray}
\end{mathletters}
The PB of these constraints are listed below
\begin{mathletters}
\begin{eqnarray}
	\left\{\phi_{0}(x),\phi_{0}(z)\right\} & = &
	\left[Q^{i}(x)+Q^{i}(z)\right]\delta_{|i}(x-z) 
	\approx 0 ~. \\    
	\left\{\phi_{0}(x),\phi_{l}(z)\right\} & = &
	\phi_{0}(z)\delta_{|l}(x-z)-\phi_{5}\lambda_{,l}(z)
	\delta(x-z)\approx 0 ~. \\    
	\left\{\phi_{0}(x),\phi_{4}(z)\right\} & = &
	\phi_{5}(z)\delta(x-z) \approx  0 ~. \\    
	\left\{\phi_{0}(x),\phi_{5}(z)\right\} & = &
	\left[B^{i}(x)+B^{i}(z)\right]\delta_{|i}(x-z)
	 +M(z)\delta(x-z)  ~. \\    
	\left\{\phi_{k}(x),\phi_{l}(z)\right\} & = &
	\phi_{l}(x)\delta_{|k}(x-z) + \phi_{k}(z)\delta_{|l}(x-z)
	\approx 0 ~. \\    
	\left\{\phi_{k}(x),\phi_{4}(z)\right\} & = &
	 0 ~. \\    
	\left\{\phi_{k}(x),\phi_{5}(z)\right\} & = &
	\phi_{5}(x)\delta_{|k}(x-z)
	-\frac{\partial\phi_{5}}{\partial\lambda}\lambda_{|k}\delta(x-z)
	 ~. \\    
	\left\{\phi_{4}(x),\phi_{4}(z)\right\} & = &
	 0 ~. \\    
	\left\{\phi_{4}(x),\phi_{5}(z)\right\} & = &
	-\frac{\partial\phi_{5}}{\partial\lambda}\delta(x-z) ~. \\    
	\left\{\phi_{5}(x),\phi_{5}(z)\right\} & = &
	\left[F^{i}(x)+F^{i}(z)\right]\delta_{|i}(x-z) ~.    
\label{ConsPB}
\end{eqnarray}
\end{mathletters}
Where the shorthanded expressions are
\begin{mathletters}
\begin{eqnarray}
    \frac{\partial\phi_{5}}{\partial\lambda} & = &
    \frac{8\pi G}{\sqrt{h}}\left[P\Theta(\Psi - \lambda I)^{-3}\Theta P 
    \right]
    \label{phi5,lambda} \\
    K_{ij} & = & -\frac{8\pi G}{\sqrt{h}}P(\Psi-\lambda I)^{-1}y_{|ij} \\
    Q^{i} & = & h^{ij}\phi_{j} + 2\left[(Kh^{ij}-K^{ij})_{|j} 
	-\frac{8\pi G}{\sqrt{h}}h^{ij}\phi_{j}\right]\phi_{5} \approx 0 \\
    B^{i} & = & \left[(Kh^{ij}-K^{ij})_{|j} -\frac{8\pi G}{\sqrt{h}}
	h^{ij}\phi_{j}\right]\frac{\partial\phi_{5}}{\partial\lambda}
	+ \left[\frac{\partial(Kh^{ij}-K^{ij})}{\partial\lambda}\right]_{|j}\phi_{5}
	\nonumber \\
	& \approx & \left[(Kh^{ij}-K^{ij})_{|j}\right]
	\frac{\partial\phi_{5}}{\partial\lambda} \\
    M & \approx & \frac{\sqrt{h}}{8\pi G}\left[\lambda\frac{\partial K}{\partial \lambda}
	- K +(R_{ij}-2K_{il}K_{j}^{l})\frac{\partial}{\partial\lambda}(Kh^{ij}-K^{ij})\right]
	\nonumber \\ 
	& & + (Kh^{ij}-K^{ij})_{|j}\left[
	\left(\frac{\partial\phi_{5}}{\partial\lambda}\right)_{|i}
	-2\frac{8\pi G}{\sqrt{h}}[P(\Psi-\lambda I)^{-1}]_{|i}(\Psi-\lambda I)^{-2}P
	 \right] \nonumber \\
	& & - \frac{8\pi G}{\sqrt{h}}
	P(\Psi-\lambda I)^{-1}[(\Psi-\lambda I)^{-1}P]_{|ij}
	\frac{\partial}{\partial\lambda}(Kh^{ij}-K^{ij})
	\\
    F^{i} & \approx & \frac{1}{3}\frac{\partial^{2}\phi_{5}}
    	{\partial \lambda^{2}}\left(Kh^{ij}-K^{ij}\right)_{|j}
    	\nonumber\\
    	& & - \left(\frac{\partial\phi_{5}}{\partial\lambda}\right)^{2}
    	\left[\left(\frac{\partial\phi_{5}}{\partial\lambda}\right)^{-1}
    	\frac{\partial}{\partial\lambda}(Kh^{ij}-K^{ij})\right]_{|j} \nonumber \\
    	& & - 2\frac{8\pi G}{\sqrt{h}} P(\Psi-\lambda I)^{-2}\left[(\Psi - \lambda I)^{-1}P
    	\right]_{|j}\frac{\partial}{\partial\lambda}(Kh^{ij}-K^{ij})
    	\label{F^i}
\end{eqnarray}
\end{mathletters}

\section{Matter Hamiltonians}
\label{app MH}
Consider here a few simple matter Lagrangians and Hamiltonians,

\noindent $\bullet$ For a cosmological constant ,
\begin{mathletters}
\begin{eqnarray}
    {\cal L}_{matter} & = & -\sqrt{-g}2\Lambda ~, \\
    T^{\alpha\beta} & = & -2\Lambda g^{\alpha\beta} ~.
\end{eqnarray}
\end{mathletters}
The corresponding energy/momentum projections are
\begin{mathletters}
\begin{eqnarray}
    T_{nn} & = & 2\Lambda ~, \\
    T_{ni} & = & 0 ~.
\end{eqnarray}
\end{mathletters}
The Hamiltonian is simply
\begin{equation}
    {\cal H}_{matter} = -{\cal L}_{matter}
    = N\sqrt{h}2\Lambda = N\sqrt{h}T_{nn}  ~.
\end{equation}

\noindent $\bullet$ For a scalar field $\Phi(x)$,
\begin{mathletters}
\begin{eqnarray}
    {\cal L}_{matter} & = & -\sqrt{-g}\left(
    \frac{1}{2}g^{\mu\nu}\partial_{\mu}\Phi\partial
    _{\nu}\Phi + V(\Phi)\right) ~, \\
    T^{\alpha\beta} & = & \left(g^{\alpha\mu}g^{\beta\nu}-
    \frac{1}{2}g^{\alpha\beta}g^{\mu\nu}\right)\partial_{\mu}
    \Phi\partial_{\nu}\Phi-g^{\alpha\beta}V(\Phi) ~.
\end{eqnarray}
\end{mathletters}
The momentum $\Pi$ conjugate to $\Phi$ is given by
\begin{equation}
    \Pi = \frac{\delta{\cal L}}{\delta \dot{\Phi}}=
    \sqrt{h}\frac{1}{N}(\dot{\Phi}-N^{i}\Phi_{,i}) ~,
    \label{Pi}
\end{equation}
and the corresponding energy/momentum projections are
\begin{mathletters}
\begin{eqnarray}
    T_{nn} & = & \frac{1}{2}\left(\frac{1}{h}\Pi^{2}+
    h^{ij}\Phi_{,i}\Phi_{,j}\right) + V ~, \\
    T_{ni} & = & \frac{1}{\sqrt{h}}\Pi\Phi_{,i} ~.
\end{eqnarray}
\end{mathletters}
The matter Hamiltonian is
\begin{equation}
    {\cal H}_{matter} = N\sqrt{h}(\frac{1}{2h}\Pi^{2}+
    \frac{1}{2}h^{ij}\Phi_{,i}\Phi_{,j} + V) + N^{i}\Pi
    \Phi_{,i} = N\sqrt{h}T_{nn} + N^{i}\sqrt{h}T_{ni} ~.
\end{equation}

\noindent $\bullet$ For a vector field $A_{\mu}(x)$,
\begin{mathletters}
\begin{eqnarray}
    {\cal L}_{matter} & = & -\frac{1}{16\pi}\sqrt{-g}
    g^{\mu\lambda}g^{\nu\sigma}F_{\mu\nu}
    F_{\lambda\sigma} ~,  \\
    T^{\alpha\beta} & = & \frac{1}{4\pi}\left(g^{\alpha\mu}
    g^{\beta\nu}-\frac{1}{4}g^{\alpha\beta}g^{\mu\nu}\right)
    g^{\lambda\sigma}F_{\mu\lambda}F_{\nu\sigma} ~.
\end{eqnarray}
\end{mathletters}
The momentum $\Pi^{\mu}$ conjugate to $A_{\mu}$ is given by
\begin{mathletters}
\begin{eqnarray}
    \Pi^{0} & = & 0 ~, \\
    \Pi^{i} & = & \frac{\sqrt{h}}{4\pi N}h^{ij}\left(
    \dot{A}_{j}-A_{0,j}-N^{k}F_{kj} ~,
    \right)
\end{eqnarray}
\end{mathletters}
and the corresponding energy/momentum projections are
\begin{mathletters}
\begin{eqnarray}
    T_{nn} & = & \frac{2\pi}{h}h^{ij}\Pi_{i}\Pi_{j}+
    \frac{1}{16\pi}h^{ij}h^{kl}F_{ik}F_{jl}  ~, \\
    T_{ni} & = & \frac{1}{\sqrt{h}}h^{kl}\Pi_{k}F_{il} ~.
\end{eqnarray}
\end{mathletters}
The Hamiltonian is
\begin{equation}
    {\cal H} = N\sqrt{h}(\frac{2\pi}{h}h^{ij}\Pi_{i}
    \Pi_{j} + \frac{1}{16\pi}h^{ij}h^{kl}F_{ik}F_{jl})
    + N^{i}\Pi^{j}F_{ij} - A_{0}\Pi^{i}_{\,,i}
    = N\sqrt{h}T_{nn}+N^{i}\sqrt{h}T_{ni}-A_{0}\Pi^{i}_{\,,i} ~.
\end{equation}
In this case the Hamiltonian picks up another Lagrange multiplier
$A_{0}$, and an additional constraint
\begin{equation}
    -\Pi^{i}_{\,,i} = \frac{1}{4\pi}\sqrt{-g}
    F^{0\nu}_{\hspace{5pt};\nu} = 0 ~.
\end{equation}

\section{The center of mass and relative coordinates}
\label{app DM}

We will try to make a canonical
transformation to the new system. We will use a global pair
${\bf Y}^{A}(t),{\bf P}_{A}(t)$ 
to describe the total momentum and
its conjugate coordinate. And, as relative coordinates we
will use the directional derivatives
$z^{A}_{\,i}(x)=y^{A}_{\,,i}(x)$ of the field $y^{A}(x)$. (This is
the analog to a discrete system, where the relative coordinates
are differences between the coordinates of the various particles
involved).

The variation of the Action with respect to $y^{A}_{\,,i}(x)$
is going to be very similar to the variation with respect to
$h_{ij}$, and therefore will resemble Einstein's equations.
The new set of canonical 'coordinates $+$ fields'
${\bf Y}^{A},{\bf P}_{A},z^{A}_{\,i}(x),\pi_{A}^{\,i}(x)$, must
obey the canonical PB
\begin{mathletters}
\begin{eqnarray}
    \left\{{\bf Y}^{A},{\bf P}_{B}\right\} & = & \delta^{A}_{B}
    ~,\label{YP} \\
    \left\{{\bf Y}^{A},\pi_{B}^{\,i}(x)\right\} & = & 0
    ~,\label{Ypi} \\
    \left\{z^{A}_{\,i}(x),{\bf P}_{B}\right\} & = & 0
    ~,\label{zP} \\
    \left\{z^{A}_{\,i}(x),\pi_{B}^{\,j}(\bar{x})\right\}
    & = & \delta^{A}_{B}\delta_{i}^{\,j}\delta(x-\bar{x})
    ~.\label{zpi}
\end{eqnarray}
\label{canPB}
\end{mathletters}
We will write the transformation from the old set of fields to the 
new set as
\begin{mathletters}
\begin{eqnarray}
    {\bf Y}^{A}(t) & = & \int d^{3}x\,f(x)y^{A}(t,x)  \\
    {\bf P}_{A}(t) & = & \int d^{3}x\,P_{A}(t,x) \\
    z^{A}_{\,i}(t,x) & = & y^{A}_{\,,i}(t,x) \\
    \pi_{A}^{\,i}(t,x) & = & \int d^{3}\bar{x}\,P_{A}
    (t,\bar{x})J^{i}(x,\bar{x})
\end{eqnarray}
\end{mathletters}
While the inverse transformation is
\begin{mathletters}
\begin{eqnarray}
    y^{A}(t,x) & = & {\bf Y}^{A}(t)+ \int d^{3}\bar{x}\,
    z^{A}_{\,i}(t,\bar{x})J^{i}(\bar{x},x) \\
    P_{A}(t,x) & = & {\bf P}_{A}(t)f(x) - \pi^{i}_{A\;,i}(t,x)
\end{eqnarray}
\end{mathletters}
The functions $f(x),J^{i}(x,\bar{x})$ are distributions over the
$V_{3}$ manifold, they do not depend on the canonical fields, and
in particular are independent of the $3$-metric. The solution to
Eq.(\ref{canPB}) put some restrictions on $f(x),J^{i}(x,\bar{x})$,
and they must fulfill the following relations
\begin{mathletters}
\begin{eqnarray}
    \int d^{3}x\,f(x) & = & 1 ~,    \\
    \int d^{3}\bar{x}\,f(\bar{x})J^{i}(x,\bar{x}) & = & 0 ~,\\
    \frac{\partial J^{i}(x,\bar{x}) }{\partial \bar{x}^{j}}
    & = & \delta^{i}_{j}\delta(x-\bar{x}) ~,\\
    \frac{\partial J^{i}(x,\bar{x}) }{\partial x^{i}}
    & = & f(x) - \delta(x-\bar{x})~.
    \label{Jf}
\end{eqnarray}
\end{mathletters}
We assume one can find such distributions and we move on to the dynamics.
We will start with the Hilbert action (\ref{action}) and do the
variation with respect to the new variables
\begin{eqnarray}
    \delta S & = & \frac{-1}{16\pi G}\int d^{4}x\,\sqrt{-g}
    (G^{\mu\nu}-8\pi G T^{\mu\nu})
    \delta g_{\mu\nu} \nonumber \\
    & = & \frac{-2}{16\pi G}\int d^{4}x\,\sqrt{-g}
    (G^{\mu\nu}-8\pi G T^{\mu\nu})
    y_{A,\mu}\delta y^{A}_{\,,\nu} \nonumber \\
    & = & \frac{2}{16\pi G}\int d^{4}x\,\left\{
    \left[\sqrt{-g}(G^{\mu 0}-8\pi G T^{\mu 0})
    y_{A,\mu}\right]_{,0}\left[\delta {\bf Y}^{A}
    + \int d^{3}\bar{x}\,\delta z^{A}_{\,i}(\bar{x})
    J^{i}(\bar{x},x)\right]\right. \nonumber \\
    & - & \frac{2}{16\pi G}\int d^{4}x\,
	\sqrt{-g}(G^{\mu i}-8\pi G T^{\mu i})
    y_{A,\mu}\delta z^{A}_{\,i}(x) ~,
    \label{daction}
\end{eqnarray}
The variation with respect to ${\bf Y}^{A}$ will lead to
the conservation of the total momentum
\begin{equation}
    \frac{-2}{16\pi G}\int d^{3}x\,
    \left[\sqrt{-g}(G^{\mu 0}-8\pi G T^{\mu 0})
    y_{A,\mu}\right] = \mu_{A} = \text{const.}
    \label{mu2}
\end{equation}
The variation with respect to $z^{A}_{\,i}(x)$ will lead to
an equation similar to Einstein's equations, but, the right
hand side does not vanish
\begin{equation}
    \sqrt{-g}\,y_{A,\alpha}[G^{\alpha i}-8\pi G T^{\alpha i}](x) =
     \int d^{3}\bar{x} J^{i}(x,\bar{x})
    \left[\sqrt{-g}(G^{\alpha 0}-8\pi G T^{\alpha 0})
    y_{A,\alpha}(\bar{x})\right]_{,0} ~.
    \label{dark2}
\end{equation}
Multiply Eq.(\ref{dark2}) by $\frac{1}{\sqrt{-g}}g^{\mu\nu}y^{A}_{\,,\nu}$
to get
\begin{equation}
    G^{\mu i}-8\pi G T^{\mu i}(x) = D^{\mu i}(x) =
    \frac{1}{\sqrt{-g}}g^{\mu\nu}y^{A}_{\,,\nu}(x) \int d^{3}\bar{x} J^{i}(x,\bar{x})
    \left[\sqrt{-g}(G^{\alpha 0}-8\pi G T^{\alpha 0})
    y_{A,\alpha}(\bar{x})\right]_{,0} ~.
    \label{dark3}
\end{equation}
An {\em{Einstein physicist}} will interpret Eq.(\ref{dark3})
as if there is some additional matter in the universe, and
may call it {\em{dark matter}}.

It is easy to reveal Eq.(\ref{RTeq}) if one takes the derivative
of Eq.(\ref{dark2}) with respect to $x^{i}$ and use (\ref{Jf}).

\section{Determinant of second class constraints PB}
\label{app detC}
We would like to calculate the determinant of $C_{mn}(x,z)$ (\ref{Cmn}).
First we will find the
eigenvalues of $C$. Take $v(x)$ to be a two components scalar function
\begin{equation}
    v(x)= \left(\begin{array}{c}
    g(x)\\ f(x) \end{array} \right) ~.
\end{equation}
The eigenvalue equation of $C$ is
\begin{equation}
    \int d^{3}z\,C(x,z)v(z)= \alpha v(x) ~.
    \label{CEV}
\end{equation}
Inserting Eq.(\ref{Cmn}) into
Eq.(\ref{CEV}) one can see that the components of $v(x)$ are
proportional, and must obey a differential equation
\begin{mathletters}
\begin{eqnarray}
    & & g = -\frac{1}{\alpha}\frac{\partial\phi_{5}}
    {\partial\lambda}f ~, \\
    & & 2F^{i}f_{|i}+F^{i}_{|i}f
    -\frac{1}{\alpha}\left(\frac{\partial\phi_{5}}
    {\partial\lambda}\right)^{2}f = \alpha f ~.
    \label{diff1}
\end{eqnarray}
\end{mathletters}
Multiplying (\ref{diff1}) by $f$ one gets
\begin{equation}
    \left(F^{i}f^{2}\right)_{|i}
    = \left[\alpha+\frac{1}{\alpha}
    \left(\frac{\partial\phi_{5}}
    {\partial\lambda}\right)^{2}\right]f^{2} ~.
    \label{diff2}
\end{equation}
Eigenvalues of a differential operator are determined by the
boundary conditions. Our boundary conditions are actually the
fact that the $3$-manifold has no boundary. Thus integrating
Eq.(\ref{diff2}) over $V_{3}$ gives us
\begin{equation}
    \int d^{3}x\left[\alpha+\frac{1}{\alpha}
    \left(\frac{\partial\phi_{5}}
    {\partial\lambda}\right)^{2}\right]f^{2}(x) =0 ~.
    \label{alpha1}
\end{equation}
Arranging Eq.(\ref{alpha1}) one gets
\begin{equation}
    \alpha^{2}= -\frac{\int d^{3}x
    \left(\frac{\partial\phi_{5}}
    {\partial\lambda}\right)^{2}f^{2}(x)}
    {\int d^{3}xf^{2}(x)} ~.
    \label{alpha2}
\end{equation}
\begin{itemize}
    \item $C_{mn}(x,z)$ is a PB matrix and therefore
    antihermitian, this causes the eigenvalues of $C$
    to be purely imaginary.
    \item One can see that the eigenvalues of $C$ are
    affected only by the off diagonal terms
    $\frac{\partial\phi_{5}}{\partial\lambda}$, not
    by $F^{i}$.
\end{itemize}
The structure of $\alpha^{2}$ is very simple. Define the
probability density
\begin{equation}
        \bar{f}(x) \equiv \frac{f^{2}(x)}
        {\int d^{3}xf^{2}(x)} ~,
\end{equation}
one sees that any eigenvalue of $C$ is simply the expectation
value of $\left(\frac{\partial\phi_{5}}
{\partial\lambda}\right)^{2}$ with respect to some probability
distribution $\bar{f}$
\begin{equation}
       \alpha^{2}_{\bar{f}}= -\left\langle\left(\frac{\partial\phi_{5}}
       {\partial\lambda}\right)^{2}\right\rangle_{\bar{f}} ~.
       \label{alpha3}
\end{equation}
For each $\bar{f}$ one finds $2$ complex conjugate purely
imaginary eigenvalues. The determinant of $C$ is therefore the
multiplication of all eigenvalues
\begin{equation}
    \det C =\prod_{\bar{f}}\left\langle\left(\frac{\partial\phi_{5}}
       {\partial\lambda}\right)^{2}\right\rangle_{\bar{f}}
    \label{detC}
\end{equation}
The probability density over a compact manifold can be
parameterized by the appropriate harmonics, and the product is
countable. See for example \cite{structure,S3} for the compact 
$S_{3}$ harmonics.

\end{document}